\documentclass[preprint,aps,tightenlines,draft]{revtex4}

\usepackage{amsmath,amssymb,psfig,epsf}

\begin{document}
\def\eqn#1{Eq.$\,$#1}
\def\mb#1{\setbox0=\hbox{$#1$}\kern-.025em\copy0\kern-\wd0
\kern-0.05em\copy0\kern-\wd0\kern-.025em\raise.0233em\box0}
\preprint{}

\title{Hamiltonian and Brownian systems with long-range interactions: II. Kinetic equations and stability analysis}
\author{Pierre-Henri Chavanis}
\affiliation{Laboratoire de Physique Th\'eorique (UMR 5152 du CNRS), Universit\'e
Paul Sabatier\\ 118 route de Narbonne, 31062 Toulouse Cedex 4,
France\\ (chavanis{@}irsamc.ups-tlse.fr)}

\begin{abstract}

\vskip 0.5cm

We discuss the kinetic theory of systems with long-range
interactions. We contrast the microcanonical description of an
isolated Hamiltonian system described by the Liouville equation from
the canonical description of a stochastically forced Brownian system
described by the Fokker-Planck equation. We show that the mean-field
approximation is exact in a proper thermodynamic limit. For
$N\rightarrow +\infty$, a Hamiltonian system is described by the
Vlasov equation. In this collisionless regime, coherent structures can
emerge from a process of violent relaxation. These metaequilibrium
states, or quasi-stationary states (QSS), are stable stationary
solutions of the Vlasov equation. To order $1/N$, the collision term
of a homogeneous system has the form of the Lenard-Balescu
operator. It reduces to the Landau operator when collective effects
are neglected. The statistical equilibrium state (Boltzmann) is
obtained on a collisional timescale of order $N$ or larger (when the
Lenard-Balescu operator cancels out). We also consider the stochastic
motion of a test particle in a bath of field particles and derive the
general form of the Fokker-Planck equation describing the evolution of
the velocity distribution of the test particle. The diffusion
coefficient is anisotropic and depends on the velocity of the test
particle.  For Brownian systems, in the $N\rightarrow +\infty$ limit,
the kinetic equation is a non-local Kramers equation. In the strong
friction limit $\xi\rightarrow +\infty$, or for large times $t\gg
\xi^{-1}$, it reduces to a non-local Smoluchowski equation. We give
explicit results for self-gravitating systems, two-dimensional
vortices and for the HMF model. We also introduce a generalized class
of stochastic processes and derive the corresponding generalized
Fokker-Planck equations. We discuss how a notion of generalized
thermodynamics can emerge in complex systems displaying anomalous
diffusion.

\pacs{???}
\vskip0.5cm
\noindent Keywords: long-range interactions; mean-field theory; Hamiltonian systems; Brownian systems 
\vskip0.5cm
\noindent Corresponding author:
P.H. Chavanis; e-mail: chavanis@irsamc.ups-tlse.fr; Tel:
+33-5-61558231; Fax: +33-5-61556065

\end{abstract}

\maketitle

\newpage

\section{Introduction}
\label{sec_introduction}

The statistical mechanics of systems with long-range interactions is
currently a topic of active research \cite{dauxois}. In a previous
paper \cite{paper1} (paper I), we have considered general models of
Hamiltonian and Brownian systems with long-range interactions, for
an arbitrary binary potential of interaction in $D$ dimensions, and
we have studied their equilibrium properties. In the present paper,
we discuss the kinetic equations describing the out-of-equilibrium
evolution of these systems.

So far, most works have focused on the case of {isolated} Hamiltonian
systems of particles in interaction such as the N-star problem in
astrophysics \cite{paddy,houches}, the N-vortex problem in 2D
hydrodynamics \cite{houches} and the HMF model \cite{hmf,cvb}.  For
these systems, the energy is conserved and the correct statistical
ensemble is the {\it microcanonical} ensemble. In a proper thermodynamic
limit $N\rightarrow +\infty$, the equilibrium one-body distribution
function maximizes the Boltzmann entropy at fixed mass and energy. The
``collisional'' evolution of the distribution function is described by
non-local kinetic equations such as the Landau-Poisson system (or the
orbit averaged Fokker-Planck equation) in astrophysics \cite{bt} or 
the kinetic equations derived by Dubin \& O'Neil \cite{dubin} and
Chavanis \cite{kin} for non-neutral plasmas and 2D point
vortices. These equations increase the Boltzmann entropy $S_B$ at
fixed mass $M$ (or circulation $\Gamma$) and energy $E$. For stellar
systems, the relaxation time (Chandrasekhar time) scales as
$t_{relax}\sim {N\over \ln N}t_{D}$, where $t_{D}$ is the dynamical
time \cite{bt}. For the point vortex gas, the evolution is due to a
condition of resonance which can be satisfied only if the profile of
angular velocity is non-monotonic
\cite{dubin,kin}. When the profile of angular velocity is monotonic,
the distribution is stationary on a timescale of at least $N t_D$
and it is not clear whether the system truly relaxes towards
statistical equilibrium for longer times. The kinetic theory of the
HMF model is also complicated \cite{bouchet,bd,cvb} and seems to
indicate a relaxation time scaling as $t_{relax}\sim N^{1.7}t_{D}$
\cite{yamaguchi}. In the ``collisionless" regime, valid for
sufficiently ``short'' times $t\ll t_{relax}$, the above-mentioned
kinetic equations reduce to the Vlasov-Poisson and 2D Euler-Poisson
systems. Since the relaxation time $t_{relax}$ increases rapidly
(algebraically) with the number of particles, the Vlasov regime can
be extremely long in practice (e.g., in astrophysics).  The
Vlasov-Poisson and 2D Euler-Poisson systems can undergo a phenomenon
of collisionless relaxation (called violent relaxation in
astrophysics) towards a metaequilibrium state on the coarse-grained
scale \cite{lb,miller,rs,csr}.  Since the relaxation time depends on
$N$, the limits $N\rightarrow +\infty$, $t\rightarrow +\infty$
(metaequilibrium) and $t\rightarrow +\infty$, $N\rightarrow +\infty$
(statistical equilibrium) differ. The metaequilibrium state is
described by non standard distribution functions which are
nonlinearly dynamically stable stationary solutions of the Vlasov
equation. In some cases, they maximize a H-function at fixed mass
and energy \cite{thlb,gt}. This maximization problem provides a
refined criterion of nonlinear dynamical stability for the
Vlasov-Poisson and 2D Euler-Poisson systems
\cite{ipser,holm,ellis,grand,cstsallis,yamaguchi,cvb}. 
The statistical equilibrium state, reached for longer times, is
described by the Boltzmann distribution.

Since statistical ensembles are not equivalent for systems with
long-range interactions, it is of interest, at a conceptual level, to
introduce a canonical model of particles with long-range
interactions. In that respect, we can consider a system of Brownian
particles described by $N$-coupled stochastic equations involving a
friction and a random force in addition to long-range forces. These
particles are in contact with a thermal bath that imposes the
temperature $T$. For these systems, the correct statistical ensemble
is the {\it canonical} ensemble.  In a proper thermodynamic limit
$N\rightarrow +\infty$, the equilibrium one-body distribution function
minimizes the Boltzmann free energy $F_B=E-TS_B$ at fixed mass $M$ and
temperature $T$. The evolution of the distribution function of these
Brownian particles is described by non-local Fokker-Planck equations
which are the canonical counterpart of the kinetic equations governing
the evolution of Hamiltonian systems.  For example, a gas of
self-gravitating Brownian particles \cite{crs} is governed by the
Kramers-Poisson and Smoluchowski-Poisson systems.  These equations
decrease the Boltzmann free energy at fixed mass and
temperature. Self-gravitating Brownian particles can experience an
``isothermal collapse'' \cite{aa}, which is the canonical version of
the ``gravothermal catastrophe'' \cite{lbw} experienced by globular
clusters. A Brownian model has also been introduced in the case of a
cosinusoidal potential of interaction in $d=1$ \cite{cvb}. This is the
canonical counterpart of the microcanonical HMF model. It could be
called the BMF (Brownian Mean Field) model.

In this paper, we compare these two descriptions (Hamiltonian and
Brownian) and study their out-of-equilibrium properties. In Sec.
\ref{sec_kin}, we discuss the kinetic theory of Hamiltonian systems
with long-range interactions by adapting the results of plasma physics
to this more general context. In Sec. \ref{sec_h}, starting from the
Liouville equation, we consider a truncation of the BBGKY hierarchy in
powers of the inverse particle number $1/N$, which plays the same role
as the plasma parameter in plasma physics. For $N\rightarrow +\infty$,
the kinetic equation is the Vlasov equation. For long-range
interactions, this equation is non-local due to mean-field effects and
exhibits a ``violent relaxation'' towards a metaequilibrium state on a
few dynamical times $t_D$ (Sec. \ref{sec_vlasov}). In
Sec. \ref{sec_ls}, we study the linear dynamical stability of a
spatially homogeneous solution of the Vlasov equation and derive a
criterion of stability generalizing the Jeans criterion in
astrophysics. We also determine analytical expressions for the growth
rate and damping rates of the perturbation.  In Sec.
\ref{sec_nonlin}, we study the nonlinear dynamical stability of a
stationary solution of the Vlasov equation and compare with the Euler
equation (Sec. \ref{sec_euler}). We also show that, for a spatially
homogenous solution, the criterion of nonlinear stability coincides
with the criterion of linear stability. In the sequel of the paper, we
consider a spatially homogeneous system which is stable with respect
to the Vlasov equation and we study the time evolution of its velocity
distribution function due to finite $N$ effects (collisions). To order
$1/N$, the kinetic equation of a homogeneous system is the Landau
equation (Sec. \ref{sec_landau}) when collective effects are ignored
and the Lenard-Balescu equation (Sec. \ref{sec_bal}) when collective
effects are properly accounted for. For 2D and 3D systems, these
equations converge towards the statistical equilibrium state
(Maxwellian distribution) on a relaxation time $t_{relax}\sim N t_{D}$
(for gravitational systems, the relaxation time is $(N/\ln N) t_{D}$
due to logarithmic divergences). For one-dimensional systems, the
Lenard-Balescu operator cancels out so that the relaxation, due to
three-body (or higher) correlations, is longer than $Nt_{D}$ (a
similar result is obtained for the point vortex gas in two dimensions
\cite{dubin,kin}). In Sec.
\ref{sec_tp}, we consider the relaxation of a test particle in a
bath of field particles and derive the general form of the
Fokker-Planck equation. The diffusion coefficient is anisotropic and
depends on the velocity.  This is responsible for anomalous
diffusion and for a slow relaxation of the high velocity tail of the
distribution \cite{bd,cl}. We provide various explicit expressions of the
diffusion coefficient and friction force for a thermal bath with
Maxwellian distribution function (subsections \ref{sec_isob} and
\ref{sec_laa}) and for one dimensional systems with an arbitrary
distribution of the bath (subsection \ref{sec_dun}). In Sec.
\ref{sec_tc}, we study the temporal correlation function of the
force and show that each mode decreases exponentially rapidly with a
decay rate which coincides with the damping rate derived in the
linear dynamical stability analysis of the Vlasov equation (Sec.
\ref{sec_ls}). In Sec. \ref{sec_ge}, we consider the time evolution
of the spatial correlation function of the particles in the linear
regime and compare with the equilibrium results obtained in Paper I.
In Sec. \ref{sec_gt}, we develop the kinetic theory of Brownian
systems with long-range interactions. Starting from the N-body
Fokker-Planck equation and using a mean-field approximation valid at
the thermodynamic limit $N\rightarrow +\infty$, we derive a
non-local Kramers equation. In the strong friction limit
$\xi\rightarrow +\infty$, or for large times $t\gg\xi^{-1}$, it
reduces to a non-local Smoluchowski equation. In Sec. \ref{sec_cw},
we study the evolution of the spatial correlation function for a
Brownian system in the linear regime. Finally, in Sec.
\ref{sec_gen}, we introduce a generalized class of stochastic
processes and derive the corresponding generalized Fokker-Planck
equations. We show that they display anomalous diffusion and that
they are associated with a notion of generalized thermodynamics in
$\mu$-space. 

One interest of our general study is to present a
unified  description of systems with long-range interactions
(Hamiltonian, Brownian, fluids,...) and to see how the results
depend on the form of the potential of interaction and on the
dimension of space $d$. Explicit results are given for gravitational
systems, two-dimensional vortices and for the HMF model. Thus, our
study shows the analogies and differences between these 
systems by placing them into a more general perspective.

\section{Kinetic theory of Hamiltonian systems} \label{sec_kin}

\subsection{The BBGKY hierarchy} \label{sec_h}

We wish to develop a kinetic theory of Hamiltonian systems with
long-range interactions described by the $N$-body equations (I-1)
of paper I in order to obtain the evolution of the one-body
distribution function $f({\bf r},{\bf v},t)=NmP_1({\bf r},{\bf
v},t)$. We will see that many results of plasma physics developed for
the Coulombian potential can be extended to a more general context. We
shall discuss how these results depend on the dimension of space and
on the form of the potential of interaction. We shall also discuss how
the results are affected by the existence of a critical point in the
case of {\it attractive} potentials (see Paper I).

Starting from the Liouville equation (I-2), it is simple to
construct the BBGKY hierarchy of equations for the reduced
distribution functions,
\begin{equation}
\label{h1} {\partial P_{j}\over\partial t}+\sum_{i=1}^{j}{\bf v}_{i}{\partial
P_{j}\over\partial {\bf r}_{i}}+\sum_{i=1}^{j}\sum_{k=1,k\neq i}^{j} {\bf F}(k\rightarrow i){\partial P_{j}\over \partial {\bf v}_{i}}+(N-j)\sum_{i=1}^{j}\int d^{D}{\bf x}_{j+1}{\bf F}(j+1\rightarrow i){\partial P_{j+1}\over\partial {\bf v}_{i}}=0.
\end{equation}
The first two equations of this
hierarchy are
\begin{equation}
\label{h2} {\partial P_{1}\over\partial t}+{\bf v}_{1}{\partial P_{1}\over\partial {\bf r}_{1}}+(N-1){\partial \over\partial {\bf v}_{1}}\int {\bf F}(2\rightarrow 1)P_{2}({\bf x}_{1},{\bf x}_{2},t)d^{D}{\bf x}_{2}=0,
\end{equation}
\begin{equation}
\label{h3}
{\partial P_{2}\over\partial t}+{\bf v}_{1}{\partial P_{2}\over\partial {\bf r}_{1}}+{\bf F}(2\rightarrow 1){\partial P_{2}\over\partial {\bf v}_{1}}+(N-2){\partial\over\partial {\bf v}_{1}}\int {\bf F}(3\rightarrow 1)P_{3}({\bf x}_{1},{\bf x}_{2},{\bf x}_{3},t)d^{D}{\bf x}_{3}+(1\leftrightarrow 2)=0,
\end{equation}
where ${\bf x}=({\bf r},{\bf v})$. Introducing the decomposition
(I-14)-(I-15), corresponding to the first terms of the Mayer
expansion in plasma physics, they can be rewritten
\begin{eqnarray}
{\partial P_{1}\over\partial t}+{\bf v}_{1}{\partial P_{1}\over\partial {\bf r}_{1}}+N{\partial P_{1}\over \partial {\bf v}_{1}}\int {\bf F}(2\rightarrow 1)P_{1}({\bf x}_{2},t)d^{D}{\bf x}_{2}\nonumber\\
+N{\partial \over\partial {\bf v}_{1}}\int {\bf F}(2\rightarrow 1)P'_{2}({\bf x}_{1},{\bf x}_{2},t)d^{D}{\bf x}_{2}=0,
\label{h4}
\end{eqnarray}
\begin{eqnarray}
\label{h5} {\partial P_{2}'\over\partial t}+{\bf v}_{1}{\partial P_{2}'\over\partial {\bf r}_{1}}+{\bf F}(2\rightarrow 1){\partial P_{2}'\over\partial {\bf v}_{1}}+{\bf F}(2\rightarrow 1)P_{1}({\bf x}_{2},t){\partial P_{1}\over\partial {\bf v}_{1}}({\bf x}_{1},t)\nonumber\\
+N{\partial\over\partial {\bf v}_{1}}\int {\bf F}(3\rightarrow 1)P_{2}'({\bf x}_{1},{\bf x}_{2},t)P_{1}({\bf x}_{3},t)d^{D}{\bf x}_{3}\nonumber\\
+N{\partial\over\partial {\bf v}_{1}}\int {\bf F}(3\rightarrow 1)P_{2}'({\bf x}_{2},{\bf x}_{3},t)P_{1}({\bf x}_{1},t)d^{D}{\bf x}_{3}\nonumber\\
+N{\partial\over\partial {\bf v}_{1}}\int {\bf F}(3\rightarrow 1)P_{3}'({\bf x}_{1},{\bf x}_{2},{\bf x}_{3},t)d^{D}{\bf x}_{3}+(1\leftrightarrow 2)=0,
\end{eqnarray}
where we have taken $N-1\simeq N$ and $N-2\simeq N$ for $N\gg 1$. We
shall now consider the thermodynamic limit defined in Sec. II.B of
paper I.  For example, we can consider $N\rightarrow +\infty$ in such
a way that the interaction potential (coupling constant) scales as
$u_*\sim 1/N$, while $\beta\sim 1$, $E/N\sim 1$ and $V\sim R^{D}\sim
1$. In that limit, the cumulant distribution functions $P_j'$ scale as
$1/N^{j-1}$. We can therefore consider an expansion of the correlation
functions in powers of the inverse particle number $1/N$. This small
parameter is the counterpart of the ``plasma parameter'' in plasma
physics. Therefore, the methods of plasma physics can be applied in
the present context with a different perspective.

\subsection{Vlasov equation and violent relaxation} \label{sec_vlasov}

For $N\rightarrow +\infty$, we get
\begin{equation}
\label{vlasov0}
P_{2}({\bf x}_{1},{\bf x}_{2},t)=P_{1}({\bf x}_{1},t)P_{1}({\bf x}_{2},t)+O(1/N)
\end{equation}
so that the mean-field approximation is exact in a proper
thermodynamic limit. In that case, the first equation of the BBGKY
hierarchy reduces to
\begin{equation}
\label{vlasov1}
{\partial f\over\partial t}+{\bf v}\cdot {\partial f\over\partial {\bf r}}+\langle {\bf F}\rangle \cdot {\partial f\over \partial {\bf v}}=0,
\end{equation}
where we have introduced  the mean-field force produced by the
particles,
\begin{equation}
\label{vlasov2}
\langle {\bf F}\rangle({\bf r},t)=\int {\bf F}(1\rightarrow 0) n({\bf r}_{1},t)d^{D}{\bf r}_{1}.
\end{equation}
The Vlasov equation (\ref{vlasov1}) expresses the conservation of the
distribution function in $\mu$-space in the absence of
``collisions''. In the case of long-range interactions, the Vlasov
equation is coupled to Eq. (\ref{vlasov2}). This coupling creates a
complicated mixing process in phase space leading to a {\it
metaequilibrium} state on a very short timescale, of the order of the
dynamical time $t_D$.  This process is called violent relaxation in
astrophysics \cite{lb} or chaotic mixing. It explains how a
collisionless system can reach a quasi-equilibrium state (on a
coarse-grained scale) as a result of phase mixing driven by long-range
interactions. This metaequilibrium state, or quasi-stationary state
(QSS), is a particular stationary solution of the Vlasov
equation. Since it results from a turbulent mixing, it is particularly
robust and has therefore nonlinear stability properties with respect
to the collisionless dynamics. For a given initial condition,
Lynden-Bell \cite{lb} has tried to predict the metaequilibrium state
reached by the system by resorting to a new type of statistical
mechanics taking into account the particularities of the collisionless
dynamics (Casimirs)
\cite{super}. Unfortunately, his theory does not always give the
good result (i.e., the distribution that is actually reached by the
system) because relaxation is incomplete in general so that the
system can be trapped in a stationary solution of the Vlasov
equation which is not the most mixed state. This concept of {\it
incomplete relaxation} \cite{lb} explains why galaxies are more
confined than predicted by Lynden-Bell's statistical mechanics. In
some cases, the metaequilibrium state turns out to maximize a
H-function \cite{thlb}:
\begin{equation}
\label{vlasov3} H[f]=-\int C(f)d^{D}{\bf r}d^{D}{\bf v},
\end{equation}
where $C$ is convex, at fixed mass $M=\int \rho d^D{\bf r}$ and
energy $E={1\over 2}\int f v^2 d^D{\bf r}d^D{\bf v}+{1\over 2}\int
\rho\Phi d^D{\bf r}$. In the context of violent relaxation, Tsallis
functional $H_{q}=-{1\over q-1}\int (f^{q}-f)d^{D}{\bf r}d^{D}{\bf
v}$ \cite{tsallis} is a particular H-function \cite{grand,gt,super,cstsallis},
not an entropy, which occasionally, but not systematically, gives a
good fit of the metaequilibrium state. Its maximization at fixed
mass and energy leads to a particular class of nonlinearly
dynamically stable stationary solutions of the Vlasov equation
called stellar polytropes in astrophysics \cite{cstsallis}. The same
interpretation holds in 2D hydrodynamics \cite{ellis,hydro} and for the
HMF model \cite{cvb,yamaguchi}. We refer to \cite{cstsallis,super}
for a more thorough discussion of these concepts.

\subsection{Linear dynamical stability} \label{sec_ls}

We consider here the linear dynamical stability of a spatially
homogeneous system with respect to the Vlasov equation
(\ref{vlasov1})-(\ref{vlasov2}). Following the standard derivation
of the stability criterion and writing $\delta f\sim {\rm
exp}\lbrack i({\bf k}\cdot {\bf r}-\omega t)\rbrack$ with
$\omega=\omega_{r}+i\lambda$, the dispersion relation reads
\cite{cvb}:
\begin{eqnarray}
\label{ls1} \epsilon({\bf k},\omega)\equiv 1-(2\pi)^{D}\hat{u}({\bf k})\int {{\bf k}\cdot {\partial f\over\partial {\bf v}}\over {\bf k}\cdot {\bf v}-\omega}d^{D}{\bf v}=0
\end{eqnarray}
where $f({\bf v})$ is the unperturbed distribution function and
$\epsilon({\bf k},\omega)$ is the dielectric function. The
integration has to be performed by using the Landau contour
\cite{ichimaru}. Note first that in the case $f({\bf
v})=\rho\delta({\bf v})$, after an integration by parts, we obtain
the explicit relation
\begin{eqnarray}
\label{ls1new} \omega^{2}=(2\pi)^{D}\hat{u}(k)k^{2}\rho.
\end{eqnarray}
On the other hand, for the Maxwellian distribution function
with uniform density
\begin{equation}
\label{maxraj} f({\bf v})=\biggl ({\beta m\over 2\pi}\biggr
)^{D/2}\rho \ e^{-\beta m {v^{2}\over 2}},
\end{equation}
the dielectric function can be expressed as
\begin{equation}
\label{ls2} \epsilon({\bf k},\omega)=1+(2\pi)^{D}\hat{u}(k)\beta m\rho  W(\sqrt{\beta m}{\omega\over k})
\end{equation}
where
\begin{equation}
\label{ls3} W(z)={1\over\sqrt{2\pi}}\int_{-\infty}^{+\infty}{x\over x-z}e^{-{x^{2}\over 2}}dx,
\end{equation}
is the $W$-function of plasma physics \cite{ichimaru}. Explicitly,
\begin{equation}
\label{ls4} W(z)=1-z e^{-{z^{2}\over 2}}\int_{0}^{z}e^{y^{2}\over 2}dy+i\sqrt{\pi\over 2}z e^{-{z^{2}\over 2}}.
\end{equation}
The dispersion relation can therefore be written
\begin{equation}
\label{ls5} 1+(2\pi)^{D}\hat{u}(k)\beta m\rho  W(\sqrt{\beta m}{\omega\over k})=0.
\end{equation}
We look for solutions of the form $\omega=i\lambda$ where $\lambda$ is real.
Then, we get
\begin{equation}
\label{ls6} \eta(k)=G\biggl (\sqrt{\beta m\over 2}{\lambda\over k}\biggr )
\end{equation}
where we have set $\eta(k)=-(2\pi)^{D}\hat{u}(k)\beta m\rho$ and defined the $G$-function
\begin{equation}
\label{ls7} G(x)={1\over 1-\sqrt{\pi}x e^{x^{2}}{\rm erfc}(x)}.
\end{equation}
For $x\rightarrow 0$, $G(x)=1+\sqrt{\pi}x+...$. For $x\rightarrow
+\infty$, $G(x)= 2x^{2}(1+{3\over 2x^{2}}+...)$.  For $x\rightarrow
-\infty$, $G(x)\sim -{1\over 2\sqrt{\pi}x}e^{-x^{2}}$. We note that
this function is always positive. Therefore, there exists solutions
of the form $\omega=i\lambda$ with $\lambda$ real only if
$\hat{u}(k)<0$, i.e. for attracting potentials. For neutral plasmas
with Maxwellian distribution, the pulsation $\omega_{r}$ of the
perturbation can never vanish.

The case of neutral stability $\omega=0$ corresponds to
$\eta(k)=1$. The case of instability $(\lambda>0)$ corresponds to
$\eta(k)>1$. In that case, the perturbation grows exponentially as
$\delta f\sim e^{\lambda t}$. The criterion (\ref{ls6}) can be
 written explicitly
\begin{equation}
\label{ls8}1-\eta(k)\biggl \lbrace 1-\sqrt{\pi m\over 2T}{\lambda\over k}e^{{m\lambda^{2}\over 2Tk^{2}}}{\rm erfc}\biggl( \sqrt{\beta m\over 2}{\lambda\over k}\biggr )\biggr\rbrace=0,
\end{equation}
where $\lambda$ is the growth rate. The case of stability $(\lambda<0)$ corresponds to $\eta(k)<1$. In that case, the perturbation decays exponentially as
$\delta f\sim e^{-\gamma t}$. The damping rate $\gamma=-\lambda$ is given by
\begin{equation}
\label{ls9} \eta(k)=F\biggl (\sqrt{\beta m\over 2}{\gamma\over k}\biggr )
\end{equation}
where we have defined the $F$-function
\begin{equation}
\label{ls10} F(x)={1\over 1+\sqrt{\pi}x e^{x^{2}}{\rm erfc}(-x)},
\end{equation}
such that $F(x)=G(-x)$. For $x\rightarrow 0$,
$F(x)=1-\sqrt{\pi}x+...$. For $x\rightarrow -\infty$,
$F(x)=2x^{2}(1+{3\over 2x^{2}}+...)$.  For $x\rightarrow +\infty$,
$F(x)\sim {1\over 2\sqrt{\pi}x}e^{-x^{2}}$. The criterion (\ref{ls9})
can be written explicitly
\begin{equation}
\label{ls11}
1-\eta(k)\biggl \lbrace 1+\sqrt{\pi m\over 2T}{\gamma\over k}e^{{m\gamma^{2}\over 2Tk^{2}}}{\rm erfc}\biggl(-\sqrt{\beta m\over 2}{\gamma\over k}\biggr )\biggr\rbrace=0.
\end{equation}
To summarize, the condition of linear dynamical stability is
$\eta(k)\le 1$ or, equivalently,
\begin{equation}
\label{ls12} 1+(2\pi)^{D}\hat{u}(k)\beta m\rho\ge 0.
\end{equation}
It coincides with the criterion of thermodynamical stability (I-95)
obtained in Paper I by considering the second order variations of
entropy or the possible divergence of the spatial correlation function
in Fourier space. The extension of these results to the case of
polytropic (Tsallis) distribution functions, instead of isothermal, is
made in
\cite{cvb} for the HMF model and in \cite{cstsallis} for
self-gravitating systems. More generally, when $f=f(v)$ depends only
on the modulus of the velocity, we have $\partial f/\partial {\bf
v}=f'(v){\bf v}/v$ and the condition of linear dynamical stability
is
\begin{equation}
\label{ls13} 1-(2\pi)^{D}\hat{u}(k)\int {f'(v)\over v}d^{D}{\bf v}\ge 0.
\end{equation}
The equality corresponds to the condition of marginal stability
$\omega=0$ in Eq. (\ref{ls1}). We refer to \cite{choi,cvb} for a
more thorough discussion of these results in the case of the HMF
model.

\subsection{The Euler equation} \label{sec_euler}

It is of interest to compare the stability of a system described by
the Vlasov equation (\ref{vlasov1})-(\ref{vlasov2}) to that of a system
described by the Euler equations
\begin{eqnarray}
\label{euler1}
{\partial \rho\over\partial t}+\nabla \cdot (\rho {\bf u})=0,
\end{eqnarray}
\begin{eqnarray}
\label{euler2}
{\partial {\bf u}\over\partial t}+({\bf u}\cdot \nabla){\bf u}=-{1\over\rho}\nabla p-\nabla\Phi,
\end{eqnarray}
with a barotropic equation of state $p=p(\rho)$ and $\Phi=\rho *
u$. The energy functional of this barotropic gas is
\begin{eqnarray}
{\cal W}[\rho,{\bf u}]=\int \rho\int_{0}^{\rho}{p(\rho')\over
\rho^{'2}}d\rho' d^{D}{\bf r}+{1\over 2}\int \rho\Phi d^{D}{\bf
r}+\int \rho {{\bf u}^2\over 2}d^{D}{\bf r}. \label{enrfun}
\end{eqnarray}
This functional is conserved by the Euler equations.
Considering the linear dynamical stability of a spatially
homogeneous system with respect to the Euler equation, we find that
the dispersion relation reads \cite{cvb}:
\begin{eqnarray}
\label{euler3} \omega^{2}=c_{s}^{2}k^{2}+(2\pi)^{D}\hat{u}({k})k^{2}\rho,
\end{eqnarray}
where $c_{s}^{2}=dp/d\rho$ is the velocity of sound. Note that for
$c_{s}=0$, Eq. (\ref{euler3}) coincides with Eq. (\ref{ls1new}). The
condition of stability is
\begin{eqnarray}
\label{euler4} c_{s}^{2}+(2\pi)^{D}\hat{u}({k})\rho\ge 0.
\end{eqnarray}
In the unstable regime, $\omega$ is imaginary ($\omega=i\lambda$
with $\lambda>0$) and the perturbation grows exponentially as
$\delta\rho\sim e^{\lambda t}$. In the stable regime, $\omega$ is
real and the perturbation oscillates as $\delta\rho\sim e^{-i\omega
t}$. For an isothermal equation of state $p=\rho T/m$, the velocity
of sound is $c_{s}^{2}=T/m$ and the instability criteria (\ref{ls12}) and
(\ref{euler4}) coincide. In fact, this coincidence
is general and goes beyond the isothermal distribution as shown in
\cite{cvb,cstsallis} and in the next section. Indeed, for spatially
homogeneous systems, the condition of dynamical stability
(\ref{ls13}) for a kinetic system described by the Vlasov equation
with $f=f(v)$ is the same as the condition of dynamical stability
(\ref{euler4}) for the corresponding barotropic gas (defined in Sec.
\ref{sec_nonlin}) described by the Euler equation. Note, however,
that the evolution of the perturbation is different in the two
systems (kinetic and gaseous) \cite{cvb}.

\subsection{Nonlinear dynamical stability} \label{sec_nonlin}

It can be shown that a distribution function $f({\bf r},{\bf v})$
which maximizes a H-function (\ref{vlasov3}) at fixed mass and energy
is a nonlinearly dynamically stable stationary solution of the Vlasov
equation. This is because $H$, $E$ and $M$ are conserved by the Vlasov
equation (see
\cite{cstsallis} for a more detailed discussion and references).
Cancelling the first variations of $H$ at fixed $E$ and $M$, we get
\begin{eqnarray}
C'(f)=-\beta\epsilon-\alpha \quad \Leftrightarrow \quad f=F(\beta\epsilon+\alpha), \label{nonlin1}
\end{eqnarray}
where $F(x)=(C')^{-1}(-x)$. We note that $f=f(\epsilon)$ depends only
on the individual energy $\epsilon={v^{2}\over 2}+\Phi({\bf r})$ of
the particles and is monotonically decreasing ($f'(\epsilon)<0$
according to Eq. (\ref{nonlin1}), assuming $\beta>0$).  This implies
that the density $\rho=\int f d^{D}{\bf v}=\rho(\Phi)$ and the
pressure $p={1\over D}\int f v^{2} d^{D}{\bf v}=p(\Phi)$ are functions
of $\Phi$. Eliminating $\Phi$ between these two expressions, we find
that the equation of state is barotropic in the sense that
$p=p(\rho)$. Therefore, to each kinetic system with distribution
function $f=f(\epsilon)$, we can associate a corresponding barotropic
gas with the same equilibrium density distribution.  Now, we have
$p'(\Phi)={1\over D}\int f'(\epsilon)v^{2}d^{D}{\bf v}={1\over D}\int
{\partial f\over\partial {\bf v}}\cdot {\bf v} d^{D}{\bf v}=-\int f
d^{D}{\bf v}$ so that $p'(\Phi)=-\rho(\Phi)$ which corresponds to the
condition of hydrostatic equilibrium in a gas $\nabla
p=-\rho\nabla\Phi$. Then, we get
$p'(\rho)=p'(\Phi)/\rho'(\Phi)=-\rho(\Phi)/\rho'(\Phi)$. But,
$\rho'(\Phi)=\int f'(\epsilon)d^{D}{\bf v}=\int {\partial
f\over\partial v}/v d^{D}{\bf v}$. Therefore, the velocity of sound
$c_{s}^{2}=p'(\rho)$ can be written $c_{s}^{2}={-\rho/\int {{\partial
f\over\partial v}/ v}d^{D}{\bf v}}$. This relation remains valid if
the system is homogeneous \cite{cvb} so that finally:
\begin{eqnarray}
c_{s}^{2}={-\rho\over \int {f'(v)\over v}d^{D}{\bf v}}.
\label{nonlin2}
\end{eqnarray}
Substituting this relation in Eq. (\ref{ls13}) we see that the
instability criteria (\ref{ls13}) and (\ref{euler4}) coincide as
announced.

We now turn to the nonlinear dynamical stability problem. It can be
shown that a distribution function $f({\bf r},{\bf v})$ which
minimizes the Casimir-Energy functional $F=E-T H$ (where $T=1/\beta$
is a positive constant) at fixed mass is a nonlinearly dynamically
stable stationary solution of the Vlasov equation. Again, this is
because $F$ and $M$ are conserved by the Vlasov equation. If the
distribution function minimizes $F$ at fixed $M$ (which is similar to
a ``canonical" stability criterion in thermodynamics) then it
maximizes $H$ at fixed $E$ and $M$ (which is similar to a
``microcanonical" stability criterion in thermodynamics)
\cite{cstsallis}. However, the reciprocal is wrong in general so that
the ``canonical" criterion $\lbrace {\rm min} \ F \ |\ M \ \rbrace$ is
less refined than the ``microcanonical" criterion $\lbrace {\rm max} \
H \ |\ M, E \ \rbrace$, and it just provides a {\it sufficient} condition of
nonlinear dynamical stability. When the two criteria do not coincide,
this is similar to a situation of ``ensemble inequivalence" in
thermodynamics.  Such ``inequivalence" is observed for the nonlinear
dynamical stability of self-gravitating systems such as stellar
polytropes and is related to the Antonov first law
\cite{grand,cstsallis}. However, for spatially homogeneous systems, 
the two criteria are equivalent in general \cite{cvb} and we shall
use here the simpler ``canonical" criterion. To minimize $F[f]$ at
fixed mass, we first minimize $F[f]$ at fixed density profile
$\rho({\bf r})$. This yields an optimal distribution $f_{*}({\bf
r},{\bf v})$ determined by $C'(f_{*})=-\beta {v^{2}\over
2}+\lambda({\bf r})$ where $\lambda({\bf r})$ can be related to the
density, using $\rho=\int f_{*}d^{D}{\bf v}$. Then, we minimize the
functional $F[\rho]=F[f_{*}]$ at fixed mass. It is shown in
\cite{grand} that this functional can be written
\begin{eqnarray}
F[\rho]=\int \rho\int_{0}^{\rho}{p(\rho')\over \rho^{'2}}d\rho' d^{D}{\bf r}+{1\over 2}\int \rho\Phi d^{D}{\bf r},
\label{nonlin3}
\end{eqnarray}
where $p(\rho)$ is the equation of state of the corresponding
barotropic gas defined above.  As observed in
\cite{grand,cstsallis,cvb}, the functional (\ref{nonlin3}) coincides
with the energy functional (\ref{enrfun}) of a barotropic gas
described by the Euler equations with ${\bf u}={\bf 0}$. Now, a
distribution $\rho({\bf r})$ which minimizes the energy functional
(\ref{enrfun}) at fixed mass is a nonlinearly dynamically stable
stationary solution of the Euler equations
(\ref{euler1})-(\ref{euler2}). This is because ${\cal W}[\rho,{\bf
u}]$ is conserved by the Euler equations. Since $F[\rho] $ and ${\cal
W}[\rho,{\bf 0}]$ coincide, the condition of nonlinear dynamical
stability for a homogeneous system described by the Vlasov equation is
the same as the condition of nonlinear dynamical stability for the
corresponding barotropic gas with respect to the Euler equation. This
is the version of the nonlinear Antonov first law for homogeneous
systems \cite{cvb}. The second variations of Eq. (\ref{nonlin3}) are
\begin{eqnarray}
\delta^{2}F=\int {p'(\rho)\over 2\rho}(\delta\rho)^{2}d^{D}{\bf r}+{1\over 2}\int \delta\rho\delta\Phi d^{D}{\bf r},
\label{nonlin4}
\end{eqnarray}
which must be positive for nonlinear dynamical stability. Now,
repeating the same steps as in Sec. IV.D  of Paper I by simply
replacing $T/m$ by $c_{s}^{2}=p'(\rho)$, since $\rho$ is constant,
we find that the condition of nonlinear dynamical stability is
\begin{eqnarray}
\label{nonlin5} c_{s}^{2}+(2\pi)^{D}\hat{u}({k})\rho\ge 0,
\end{eqnarray}
for all $k$. Using Eq. (\ref{nonlin2}), this can also be expressed
as
\begin{equation}
\label{ls13nonlin} 1-(2\pi)^{D}\hat{u}(k)\int {f'(v)\over
v}d^{D}{\bf v}\ge 0,
\end{equation}
for all $k$. This generalizes the criterion of nonlinear dynamical
stability given in \cite{yamaguchi} for the HMF model. Our approach
also gives an interpretation of this criterion in terms of a
condition on the velocity of sound in a barotropic gas with the same
equation of state as the original kinetic system \cite{cvb}.
Indeed, we must have
\begin{eqnarray}
\label{nonlin6} c_{s}^{2}>(c_{s}^{2})_{max}\equiv
\rho\hat{v}(k)_{max}
\end{eqnarray}
for nonlinear dynamical stability. Finally, we note that the
criteria (\ref{nonlin5}) and (\ref{ls13nonlin}) are the same as
(\ref{euler4}) and (\ref{ls13}). This implies that, for homogeneous
systems, the conditions of linear and nonlinear dynamical stability
coincide.

\subsection{The Landau equation} \label{sec_landau}

The Vlasov equation is very important for systems with long-range
interactions because collisional effects become manifest on a
timescale of order of $N t_D$ or larger. For many realistic systems
(e.g., galaxies in astrophysics) the number of particles is so large
($N\sim 10^{12}$) that only the collisionless regime matters for
timescales of interest \cite{bt}. Therefore, we can consider the
limit $N\rightarrow +\infty$ leading rigorously to the Vlasov
equation. However, the limits $N\rightarrow \infty$ and
$t\rightarrow \infty$ are not interchangeable. For sufficiently long
times, collisional effects must be taken into account. This is the
case for globular clusters in astrophysics \cite{bt} which form
smaller groups of stars ($N\sim 10^{6}$) and whose age is of the
order of the relaxation time. We would like now to take the effect
of correlations between particles into account in order to describe
the ``collisional'' relaxation due to finite $N$ effects. In
particular, we would like to obtain the form of the collision term
to order $1/N$. This is the first correction to the Vlasov regime in
the $N^{-1}$ expansion of the correlation functions. It will
describe the dynamics of the system on a timescale $\sim N t_{D}$.

There are different methods to obtain a kinetic equation for the
distribution function $f({\bf r},{\bf v},t)$. One possibility is to
close the BBGKY hierarchy by neglecting the cumulant of the
three-body correlation function \cite{ichimaru}, which is of order
$P_{3}'\sim N^{-2}$ in the present context. This corresponds to the
Kirkwood approximation in plasma physics.  Another possibility is to
use the Klimontovich approach and develop a quasilinear theory (see, e.g., 
\cite{paddy} and Appendix \ref{sec_leba}). A third possibility is to
use a projection operator formalism, e.g. \cite{kandrup1}. An interest of
this approach is that it takes into account non Markovian effects
and spatial delocalization. If we neglect collective effects, the
projection operator formalism leads to a kinetic equation of the
form
\begin{eqnarray}
\label{landau1} {\partial f\over\partial t}+{\bf v}\cdot {\partial f\over\partial {\bf r}}+\langle {\bf F}\rangle\cdot {\partial f\over\partial {\bf v}}={\partial\over\partial v^{\mu}}\int_{0}^{t}d\tau\int d^{D}{\bf r}_{1}d^{D}{\bf v}_{1}{\cal F}^{\mu}(1\rightarrow 0,t)\nonumber\\
\times \biggl\lbrace {\cal F}^{\nu}(1\rightarrow 0, t-\tau){\partial\over\partial v^{\nu}}+{\cal F}^{\nu}(0\rightarrow 1, t-\tau)   {\partial\over\partial {v}_{1}^{\nu}}\biggr \rbrace {f\over m}({\bf r}_{1},{\bf v}_{1},t-\tau)f({\bf r},{\bf v},t-\tau).
\end{eqnarray}
Here, $f({\bf r},{\bf v},t)=NmP_{1}({\bf r},{\bf v},t)$ is the
distribution function, $\langle {\bf F}\rangle({\bf r},t)$ is the
(smooth) mean-field force and ${\cal F}^{\mu}(1\rightarrow
0,t)=F^{\mu}(1\rightarrow 0,t)-\langle F^{\mu}\rangle({\bf r},t)$ is
the fluctuating force created by particle $1$ (with position and
velocity ${\bf r}_{1},{\bf v}_{1}$) on particle $0$ (with ${\bf
r},{\bf v}$) at time $t$. Between $t$ and $t-\tau$, the particles
are assumed to follow the trajectories determined by the slowly
evolving mean-field $\langle {\bf F}\rangle({\bf r},t)$.  Equation
(\ref{landau1}) is a non-Markovian integrodifferential equation. We
insist on the fact that this equation is valid for an inhomogeneous
system while the kinetic equations presented in the sequel will only
apply to homogeneous systems. Unfortunately, Eq. (\ref{landau1})
remains too complicated for practical purposes and we have to
make simplifications. If we consider a spatially homogeneous system
for which the distribution function $f=f({\bf v},t)$ depends only on
the velocity \footnote{This assumes that, at each time $t$, the
distribution function $f({\bf v},t)$ is a stable stationary solution
of the Vlasov equation. Therefore, we study how the velocity
distribution of a spatially homogeneous system changes due to
``collisions'' (finite $N$ effects). This implicitly assumes that
the energy is higher than the critical energy $E_c$ at which the
homogeneous phase becomes unstable at statistical equilibrium (see
Paper I).}, and if we implement a Markovian approximation, the
foregoing equation reduces to
\begin{eqnarray}
\label{landau2} {\partial f\over\partial t}={\partial\over\partial v^{\mu}}\int_{0}^{+\infty}d\tau\int d^{D}{\bf r}_{1}d^{D}{\bf v}_{1}{F}^{\mu}(1\rightarrow 0,t)\nonumber\\
\times {F}^{\nu}(1\rightarrow 0, t-\tau)\biggl ({\partial\over\partial v^{\nu}}-{\partial\over\partial {v}_{1}^{\nu}}\biggr ){f\over m}({\bf v}_{1},t)f({\bf v},t),
\end{eqnarray}
with $F^{\mu}(1\rightarrow 0,t)=F^{\mu}({\bf r}_{1}(t)\rightarrow {\bf
r}(t))$. In our approximation,  the particles follow linear trajectories with constant velocity since $\langle {\bf F}\rangle={\bf 0}$. Then, the collision term can be simplified (see Appendix \ref{sec_lt}) and we
obtain
\begin{eqnarray}
\label{landau3} {\partial f\over\partial t}=\pi (2\pi)^{D}m{\partial\over\partial v^{\mu}}\int d^{D}{\bf v}_{1}d^{D}{\bf k}k^{\mu}k^{\nu}{\hat{u}({\bf k})^{2}}\delta\lbrack {\bf k}\cdot ({\bf v}-{\bf v}_{1})\rbrack \biggl (f_{1}{\partial f\over\partial v^{\nu}}-f{\partial f_{1}\over\partial {v}_{1}^{\nu}}\biggr ),
\end{eqnarray}
where $f=f({\bf v},t)$ and $f_{1}=f({\bf v}_{1},t)$. Introducing the
relative velocity ${\bf u}={\bf v}-{\bf v}_{1}$, this can be written more conveniently as
\begin{eqnarray}
\label{landau4} {\partial f\over\partial t}={\partial\over\partial
v^{\mu}}\int d^{D}{\bf v}_{1}K^{\mu\nu}({\bf u})  \biggl
(f_{1}{\partial f\over\partial v^{\nu}}-f{\partial
f_{1}\over\partial {v}_{1}^{\nu}}\biggr ),
\end{eqnarray}
where
\begin{eqnarray}
\label{landau5}K^{\mu\nu}=\pi (2\pi)^{D}m\int d^{D}{\bf k} k^{\mu}k^{\nu}\hat{u}(k)^{2}\delta({\bf k}\cdot {\bf u}).
\end{eqnarray}
We note that the thermodynamic limit defined in Paper I
amounts to considering that the coupling constant scales as $1/N$, all
other quantities being of order unity. Then, recalling that
$f=NmP_{1}\sim N$, we find that the collision operator in the Landau
equation scales as $1/N$ and represents therefore the first correction
to the Vlasov equation. It also results from this observation that the
 Landau equation will describe the evolution of the system on a timescale of
order  $Nt_{D}$.

Equation (\ref{landau4}) with Eq. (\ref{landau5}) is the general
form of the Landau equation. Landau derived it for an electronic
plasma from the Boltzmann equation in a weak deflexion limit, using
a linear trajectory approximation \cite{balescu}. It can also be
obtained from the Fokker-Planck equation by calculating the first
and second moments of the velocity increments induced by a
succession of two-body encounters \cite{chanstell}. The Landau
equation can be further simplified by explicitly evaluating the
tensor (\ref{landau5}).  In $D=3$ and $D=2$, we find that
\begin{eqnarray}
\label{landau7}K^{\mu\nu}=K_{D}\biggl
(\delta^{\mu\nu}-{u^{\mu}u^{\nu}\over u^{2}}\biggr ),
\end{eqnarray}
where
\begin{eqnarray}
\label{landau8}K_{3}=8\pi^{5}m{1\over
u}\int_{0}^{+\infty}k^{3}\hat{u}(k)^{2}dk, \qquad
K_{2}=8\pi^{3}m{1\over u}\int_{0}^{+\infty}k^{2}\hat{u}(k)^{2}dk.
\end{eqnarray}
The Landau equation conserves the constants of motion of the
Hamiltonian dynamics (mass, energy,...) and increases the Boltzmann
entropy (H-theorem). For $D=2$ and $D=3$, the distribution function
approaches the Maxwellian for $t\rightarrow +\infty$.  The relaxation
time scales as $t_{relax}\sim N t_D$. Therefore, the kinetic theory
justifies, at least at order $1/N$ and for homogeneous systems, the
Maxwell distribution predicted by the statistical theory. Since the
microcanonical distribution is based on an assumption of ergodicity
(see Paper I), it is not granted a priori that the system reaches
statistical equilibrium. Thus, the development of a kinetic theory is
necessary to validate (or not) the microcanonical distribution. For
inhomogeneous systems, it is not clear whether the kinetic equation
(\ref{landau1}) truly relaxes towards the mean-field Maxwell-Boltzmann
distribution due to non-Markovian effects and spatial delocalization.
Therefore, the validity of the Boltzmann distribution depends on the
dynamics and a kinetic theory is required to justify its relevance.

The collisional evolution of stellar systems (that are
inhomogeneous) is usually described by the Vlasov-Landau equation
\begin{eqnarray}
\label{landau6}  {\partial f\over\partial t}+{\bf v} \cdot {\partial
f\over\partial {\bf r}}+\langle {\bf F}\rangle\cdot {\partial
f\over\partial {\bf v}}={\partial\over\partial v^{\mu}}\int
d^{D}{\bf v}_{1}K^{\mu\nu}({\bf u}) \biggl (f_{1}{\partial
f\over\partial v^{\nu}}-f{\partial f_{1}\over\partial
{v}_{1}^{\nu}}\biggr ),
\end{eqnarray}
with now $f=f({\bf r},{\bf v},t)$ and $f_{1}=f({\bf r},{\bf
v}_{1},t)$. This equation, obtained by combining Eqs.
(\ref{vlasov1}) and (\ref{landau4}), assumes that the collisions can
be treated as local (see \cite{kandrup1} for a critical discussion
of this approximation). It has to be coupled to the Poisson
equation. For the gravitational potential in $D=3$,
\begin{eqnarray}
\label{landau9}K_{3}=2\pi mG^{2}{1\over u}\int_{0}^{+\infty}{dk\over k}.
\end{eqnarray}
This quantity exhibits a well-known logarithmic divergence at small
and large scales. The divergence at small scales is common to both
plasmas and gravitational systems. It is due to the failure of the
linear trajectory approximation. It is regularized by cutting the
integral at the Landau length, corresponding to the impact parameter
for large angle collisions.  The divergence at large scales is
specific to the gravitational case (in the plasma case, the integral
must be cut-off at the Debye length as shown by the Lenard-Balescu
treatment). It is due to the long-range nature of gravity and the
absence of shielding. This problem has been the subject of several
studies. In their stochastic analysis, Chandrasekhar \& von Neumann
\cite{neumann} argue that the integral has to be cut-off at the
interparticle distance since the distribution of the gravitational
field (a L\'evy distribution) is dominated by the contribution of
the nearest neighbor. Alternatively, most astrophysicists argue that
the integral has to be cut-off at the size of the system, of the
order of the Jeans length, which is presumably the gravitational
analogue of the Debye length. In any case, because of the
logarithmic divergence, the relaxation time scales as $t_{relax}\sim
(N/\ln N)t_{D}$ instead of $Nt_{D}$. For the gravitational or
Coulombian potential in $D=2$, the integral behaves as
$\int_{0}^{+\infty}dk/k^{2}$ and it diverges linearly for
$\lambda=2\pi/k \rightarrow +\infty$. This suggests that $K_{2}$ is
proportional to the large-scale cut-off (the Debye length in plasma
physics). However, a more precise study based on the Lenard-Balescu
equation taking into account collective effects is required to
ascertain this result.

In $D=1$, the Landau equation becomes
\begin{eqnarray}
\label{laudau10} {\partial f\over\partial t}=K{\partial\over\partial v}\int d{v}_{1} \delta(v-v_{1}) \biggl (f_{1}{\partial f\over\partial v}-f{\partial f_{1}\over\partial {v}_{1}}\biggr ),
\end{eqnarray}
where
\begin{eqnarray}
\label{laudau11}K=4\pi^{2}m\int_{0}^{+\infty}k\hat{u}(k)^{2}dk.
\end{eqnarray}
Due to the $\delta$-function, the collision term vanishes
identically. This implies that the distribution function does not
evolve on a timescale of order $N t_D$, i.e. $\partial f/\partial
t=0$. Since the Landau equation is valid at order $1/N$, the
relaxation towards statistical equilibrium will be due to non
trivial correlations which appear at higher order in the
large $N$ expansion. This implies that, in 1D, the relaxation time
is longer than $Nt_{D}$. For the HMF model, Yamaguchi {\it et al.}
\cite{yamaguchi} report a non trivial scaling $N^{1.7}$. A similar
cancellation of the collision term at order $1/N$ occurs for 2D point
vortices when the profile of angular velocity is monotonic. Indeed,
neglecting collective effects, the kinetic equation describing the
collisional evolution of an axisymmetric point vortex system is
\cite{kin}:
\begin{eqnarray}
\label{landau12}
{\partial P\over\partial t}=-{N\gamma^{2}\over 4r}{\partial\over\partial r}\int_{0}^{+\infty}r_{1}dr_{1}\delta(\Omega-\Omega_{1})\ln\biggl \lbrack 1-\biggl ({r_{<}\over r_{>}}\biggr )^{2}\biggr\rbrack \biggl ({1\over r}P_{1}{\partial P\over\partial r}-{1\over r_{1}}P{\partial P_{1}\over\partial r_{1}}\biggr ),
\end{eqnarray}
where $P=P(r,t)$, $P_{1}=P(r_{1},t)$ and $r_<$ (resp. $r_>$) is the
min (resp. max) of $r$ and $r_{1}$. The evolution is due to distant
collisions. The collision operator is due to a resonance between
vortices rotating with equal angular velocity
$\Omega(r)=\Omega(r_{1})$ and vanishes identically when there is no
resonance. Equation (\ref{landau12}), derived in Chavanis
\cite{kin} by using projection operator technics, is the vortex
analogue of the Landau equation (\ref{landau4}) for plasmas.  A more
general kinetic equation, taking into account collective effects, has
been derived by Dubin \& O'Neil \cite{dubin} from the Klimontovich
approach. It represents the analogue of the Lenard-Balescu equation
(\ref{bal1}) in plasma physics. As discussed above, the relaxation of
2D point vortices and of the HMF model toward statistical equilibrium
is not clearly understood and demands to go to higher order in the
$1/N$ expansion.

\subsection{The Lenard-Balescu equation} \label{sec_bal}

The Lenard-Balescu equation can be obtained from the first two
equations of the BBGKY hierarchy by neglecting non-trivial three-body
correlations and assuming that the two-particle correlation function
relaxes much faster than the one-particle distribution function
(Bogoliubov's hypothesis) \cite{ichimaru}. It can also be obtained
from the Klimontovich equation by developing a quasilinear theory (see
Appendix \ref{sec_leba}). For a homogeneous system, the Lenard-Balescu
equation can be written
\begin{eqnarray}
\label{bal1} {\partial f\over\partial t}=\pi (2\pi)^{D}m{\partial\over\partial v^{\mu}}\int d^{D}{\bf v}_{1}d^{D}{\bf k}k^{\mu}k^{\nu}{\hat{u}({\bf k})^{2}\over |\epsilon({\bf k},{\bf k}\cdot {\bf v})|^{2}}\delta\lbrack {\bf k}\cdot ({\bf v}-{\bf v}_{1})\rbrack \biggl (f_{1}{\partial f\over\partial v^{\nu}}-f{\partial f_{1}\over\partial {v}_{1}^{\nu}}\biggr )
\end{eqnarray}
where
\begin{eqnarray}
\label{bal2} \epsilon({\bf k},\omega)=1+(2\pi)^{D}\hat{u}({\bf k})\int {{\bf k}\cdot {\partial f\over\partial {\bf v}}\over \omega-{\bf k}\cdot {\bf v}}d^{D}{\bf v},
\end{eqnarray}
is the dielectric function. The Lenard-Balescu equation can be seen
as a generalization of the Landau equation taking into account
collective effects. The classical Landau equation is recovered in
the limit $|\epsilon({\bf k},{\bf k}\cdot {\bf v})|^{2}\simeq 1$. In
D=1, the Lenard-Balescu equation becomes
\begin{eqnarray}
\label{bal3} {\partial f\over\partial t}=2\pi^{2}m{\partial\over\partial v}\int d{v}_{1} |k| d{k}{\hat{u}({k})^{2}\over |\epsilon({k},{k}{v})|^{2}}\delta ({v}-{v}_{1})\biggl (f_{1}{\partial f\over\partial v}-f{\partial f_{1}\over\partial {v}_{1}}\biggr )
\end{eqnarray}
where
\begin{eqnarray}
\label{bal4} \epsilon({k},\omega)=1+2\pi\hat{u}({k})\int {f'({v})\over \omega/k-{v}}d{v}.
\end{eqnarray}
We see that the collision term again vanishes identically at the order
$1/N$.

\subsection{Test particle in a thermal bath: the Fokker-Planck equation} \label{sec_tp}

The Lenard-Balescu equation can also be used to describe the evolution
of a test particle in a bath of field particles at equilibrium. In
that case, we have to consider that the distribution $f_{1}$ of the
bath is {\it given}, i.e. $f_{1}({\bf v},t)=f_{0}({\bf v})$ where
$f_{0}({\bf v})$ is a stable stationary solution of the Vlasov
equation (bath distribution). This procedure  transforms the
integro-differential equation (\ref{bal1}) into a differential
 equation for the density probability $P({\bf v},t)$  of finding the test particle with velocity ${\bf v}$ at time $t$. It reads
\begin{eqnarray}
\label{tp1} {\partial P\over\partial t}=\pi (2\pi)^{D}m{\partial\over\partial v^{\mu}}\int d^{D}{\bf v}_{1}d^{D}{\bf k}k^{\mu}k^{\nu}{\hat{u}({\bf k})^{2}\over |\epsilon({\bf k},{\bf k}\cdot {\bf v})|^{2}}\delta\lbrack {\bf k}\cdot ({\bf v}-{\bf v}_{1})\rbrack \biggl ({\partial \over\partial v^{\nu}}-{\partial \over\partial {v}_{1}^{\nu}}\biggr )f_{0}({\bf v}_{1})P({\bf v},t),\nonumber\\
\end{eqnarray}
where
\begin{eqnarray}
\label{tp2} \epsilon({\bf k},\omega)=1+(2\pi)^{D}\hat{u}({\bf k})\int {{\bf k}\cdot {\partial f_{0}\over\partial {\bf v}}\over \omega-{\bf k}\cdot {\bf v}}d^{D}{\bf v}.
\end{eqnarray}
Equation (\ref{tp1}) can be written in the form of a Fokker-Planck equation
\begin{equation}
\label{tp3}{\partial P\over\partial t}={\partial\over\partial v^{\mu}}\biggl (D^{\mu\nu}{\partial P\over\partial v^{\nu}}+P\eta^{\mu}\biggr ),
\end{equation}
which involves a diffusion term
\begin{equation}
\label{tp4} D^{\mu\nu}=\pi (2\pi)^{D}m\int d^{D}{\bf v}_{1}d^{D}{\bf k}k^{\mu}k^{\nu}{\hat{u}({\bf k})^{2}\over |\epsilon({\bf k},{\bf k}\cdot {\bf v})|^{2}}\delta\lbrack {\bf k}\cdot ({\bf v}-{\bf v}_{1})\rbrack f_{0}({\bf v}_{1})
\end{equation}
and a friction term
\begin{equation} \eta^{\mu}=-\pi (2\pi)^{D}m\int d^{D}{\bf v}_{1}d^{D}{\bf k}k^{\mu}k^{\nu}{\hat{u}({\bf k})^{2}\over |\epsilon({\bf k},{\bf k}\cdot {\bf v})|^{2}}\delta\lbrack {\bf k}\cdot ({\bf v}-{\bf v}_{1})\rbrack {\partial f_{0}\over\partial {v}_{1}^{\nu}}({\bf v}_{1}).
\label{tp5}
\end{equation}
The diffusion coefficient is due to the fluctuations of the force and is given by the Kubo formula
\begin{eqnarray}
\label{tp8} D^{\mu\nu}=\int_{0}^{+\infty} \langle F^{\mu}(0)F^{\nu}(t)\rangle dt.
\end{eqnarray}
The dynamical friction results from a polarization process and  can be
explicitly calculated by developing a linear response theory. In the
present context, the coefficients of diffusion and friction depend on
the velocity.  Hence, it is more proper to write Eq. (\ref{tp3}) in a
form which is fully consistent with the general Fokker-Planck equation
\begin{equation}
\label{tp6} {\partial P\over\partial t}={1\over 2}{\partial^{2}\over\partial v^{\mu}\partial v^{\nu}}\biggl (P{\langle \Delta v^{\mu} \Delta v^{\nu}\rangle\over \Delta t}\biggr )-{\partial\over\partial v^{\mu}}\biggl (P{\langle \Delta v^{\mu}\rangle\over \Delta t}\biggr ),
\end{equation}
with
\begin{equation}
\label{tp7}{\langle \Delta v^{\mu} \Delta v^{\nu}\rangle\over \Delta
t}=2D^{\mu\nu}, \qquad {\langle \Delta v^{\mu}\rangle\over \Delta
t}={\partial D^{\mu\nu}\over\partial
v^{\nu}}-\eta^{\nu}=-2\eta^{\nu},
\end{equation}
where the last equality is obtained from   Eqs. (\ref{tp4}) and
(\ref{tp5}) by using an integration by parts. We refer to Ichimaru
\cite{ichimaru} for a more comprehensive discussion of the test
particle approach and for the connection between the Lenard-Balescu
equation and the Fokker-Planck equation in the thermal bath
approximation. Our aim, here, is to give explicit expressions of the
Fokker-Planck equation and diffusion coefficient in particular
cases.

\subsubsection{The isothermal case} \label{sec_isob}

For the Boltzmann distribution
\begin{equation}
\label{isob1} f_{0}({\bf v}_{1})=\biggl ({\beta m\over 2\pi}\biggr )^{D/2}\rho \ e^{-\beta m {v_{1}^{2}\over 2}},
\end{equation}
corresponding to statistical equilibrium (thermal bath), we easily  find
from Eqs. (\ref{tp4}) and (\ref{tp5}) that $\eta^{\mu}=\beta m
D^{\mu\nu}v^{\nu}$. Therefore, the friction coefficient is given by a generalized Einstein
relation and the Fokker-Planck equation (\ref{tp3}) takes the
form
\begin{equation}
\label{isob2} {\partial P\over\partial t}={\partial\over\partial v^{\mu}}\biggl\lbrack D^{\mu\nu}({\bf v})\biggl ({\partial P\over\partial v^{\nu}}+\beta m P v^{\nu}\biggr )\biggr\rbrack.
\end{equation}
This is similar to the Kramers equation in Brownian theory. However,
in the present context, the diffusion coefficient $D^{\mu\nu}$ is
anisotropic and depends on the velocity ${\bf v}$ of the test
particle. We note that for $t\rightarrow +\infty$, the velocity
distribution of the test particle $P({\bf v},t)$ relaxes to the
Maxwellian distribution (\ref{isob1}) of the bath (thermalization).
Writing
\begin{equation}
\label{isob3} \delta\lbrack {\bf k}\cdot ({\bf v}-{\bf v}_{1})\rbrack=\int e^{i{\bf k}\cdot ({\bf v}-{\bf v}_{1})t}{dt\over 2\pi},
\end{equation}
we can put the diffusion coefficient (\ref{tp4}) in the form
\begin{equation}
\label{isob4} D^{\mu\nu}({\bf v})= (2\pi)^{2D}m \int_{0}^{+\infty} dt \int d^{D}{\bf k}{k^{\mu}k^{\nu}}{\hat{u}({\bf k})^{2}\over |\epsilon({\bf k},{\bf k}\cdot {\bf v})|^{2}}e^{i{\bf k}\cdot {\bf v}t}\hat{f}_{0}({\bf k}t).
\end{equation}
It will be shown in Sec. \ref{sec_tc} that this equation can be
interpreted as a Kubo formula (\ref{tp8}) where $t$ plays the role of
time. For a Maxwellian distribution, the Fourier transform
$\hat{f}_{0}({\bf k}t)$ is a Gaussian. We can thus easily integrate on
time $t$ to obtain
\begin{equation}
\label{isob5} D^{\mu\nu}({\bf v})=\pi (2\pi)^{D}m \rho \biggl ({\beta m\over 2\pi}\biggr )^{1/2}\int d^{D}{\bf k}{k^{\mu}k^{\nu}\over k}{\hat{u}({\bf k})^{2}\over |\epsilon({\bf k},{\bf k}\cdot {\bf v})|^{2}}e^{-\beta m {({\bf k}\cdot {\bf v})^{2}\over 2k^{2}}}.
\end{equation}
On the other hand, for a Maxwellian distribution, the dielectric function is given by Eq. (\ref{ls2}). Using (\ref{ls4}), we can thus rewrite the Fokker-Planck equation in the form
\begin{equation}
\label{isob8} {\partial P\over\partial t}={1\over t_{R}}{\partial\over\partial x^{\mu}}\biggl\lbrack G^{\mu\nu}({\bf x})\biggl ({\partial P\over\partial x^{\nu}}+2 P x^{\nu}\biggr )\biggr\rbrack,
\end{equation}
where
\begin{equation}
\label{isob9} G^{\mu\nu}({\bf x})=L^{D+1}\int d^{D}{\bf k}{\hat{k}^{\mu}\hat{k}^{\nu}k\eta(k)^{2}e^{-(\hat{\bf k}\cdot {\bf x})^{2}}\over \lbrack 1-\eta(k)B(\hat{\bf k}\cdot {\bf x})\rbrack^{2}+\pi \eta(k)^{2}(\hat{\bf k}\cdot {\bf x})^{2}e^{-2(\hat{\bf k}\cdot {\bf x})^{2}}}
\end{equation}
and $t_{R}$ is a relaxation timescale defined by
\begin{equation}
\label{isob10} t_{R}^{-1}={1\over (2\pi)^{D}}\biggl ({\pi\over 8D}\biggr )^{1/2}{v_{m}\over n}{1\over L^{D+1}}.
\end{equation}
In the foregoing formulae, we have set ${\bf x}=({\beta
m/2})^{1/2}{\bf v}$, $\hat{\bf k}={\bf k}/k$, $B(x)=1-2x
e^{-x^{2}}\int_{0}^{x}e^{y^{2}}dy$ and
$\eta(k)=-(2\pi)^{D}\hat{u}(k)\beta m\rho$. In addition,
$v_{m}=({D/\beta m})^{1/2}$ is the r.m.s. velocity and $L$ is a
lengthscale (size of the domain) introduced to make Eq.
(\ref{isob9}) dimensionless. The function $B(x)$ can be written
$B(x)=1-2x D(x)$ where $D(x)=e^{-x^{2}}\int_{0}^{x}e^{y^{2}}dy$ is
Dawson's integral. We note the asymptotic behaviors
$B(x)=1-2x^{2}+...$ for $x\rightarrow 0$ and $B(x)\sim -{1\over
2x^{2}}$ for $x\rightarrow +\infty$.  Equations
(\ref{isob8})-(\ref{isob10}) provide the general form of the
Fokker-Planck equation describing the evolution of a test particle
in a thermal bath. They generalize the results obtained in
\cite{bouchet,bd,cvb} for the HMF model where the potential of
interaction is truncated to one single mode and $D=1$. We note that
the relaxation time scales as $t_{R}\sim Nt_{D}$ where we have
introduced the dynamical time $t_{D}\sim L/v_{m}$ and the particle
number $N\sim nL^{D}$. As indicated previously, the Fokker-Planck
collision term scales as $1/N$ in the thermodynamic limit defined in Paper I.

We also note that in the case of the Coulombian interaction for
which $\eta(k)=-k_{D}^{2}/k^{2}$ where $k_D$ is the Debye
wavenumber, the integral over $k$ in Eq. (\ref{isob9}) is of the
form
\begin{equation}
\label{lbq1} I_{D}=\int_{0}^{+\infty}{k^{D}dk\over (B+k^{2}/k_{D}^{2})^2+C^{2}}
\end{equation}
where $B\equiv B(\hat{\bf k}\cdot {\bf x})$ and $C\equiv \sqrt{\pi}
|\hat{\bf k}\cdot {\bf x}|e^{-(\hat{\bf k}\cdot {\bf x})^{2}}$ are
independent of $k$. The integral (\ref{lbq1}) can be rewritten
$I_{D}={k_{D}^{D+1} C^{D-3\over 2}}\Phi_{D}(B/C)$ with
$\Phi_{D}(x)=\int_{0}^{+\infty}t^{D}dt/((x+t^{2})^{2}+1)$.  In the
Landau approximation, we have instead
$I_{D}^{Landau}=k_{D}^{4}\int_{0}^{+\infty}k^{D-4}dk$ which presents
the divergences mentioned previously. In $D=3$, the integral
(\ref{lbq1}) is now convergent for $k\rightarrow 0$ and, using the
dominant approximation of Chandrasekhar \cite{ichimaru}, we have
$I_{3}\sim k_{D}^{4}\ln(k_{m}/k_{D})$ where $k_{m}$ is a small scale
cut-off. Therefore, the Lenard-Balescu treatment justifies to cut the
logarithmic divergence at large scales at the Debye length
$k_{D}^{-1}$ \cite{ichimaru}. In $D=2$, we note that the integral
(\ref{lbq1}) is convergent for all values of $k$. The evaluation of
the diffusion coefficient (\ref{isob9}) in that case is left for a
future study.  Finally, the case $D=1$ will be treated in
Sec. \ref{sec_dun}.

\subsubsection{The Landau approximation} \label{sec_laa}

If we make the Landau approximation $|\epsilon({\bf k},{\bf k}\cdot {\bf v})|^{2}\simeq 1$ which amounts to neglecting collective effects, we can rewrite the Fokker-Planck equation in the form
\begin{equation}
\label{laa1} {\partial P\over\partial t}={1\over t_{R}}{\partial\over\partial x^{\mu}}\biggl\lbrack G^{\mu\nu}({\bf x})\biggl ({\partial P\over\partial x^{\nu}}+2 P x^{\nu}\biggr )\biggr\rbrack,
\end{equation}
where now
\begin{equation}
\label{laa2} G^{\mu\nu}({\bf x})=\int d^{D}\hat{\bf k}{\hat{k}^{\mu}\hat{k}^{\nu}e^{-(\hat{\bf k}\cdot {\bf x})^{2}}},
\end{equation}
\begin{equation}
\label{laa3} t_{R}^{-1}=\bigl ({\pi\over 8}\bigr )^{1/2}D^{3/2}(2\pi)^{D}{\rho m\over v_{m}^{3}}\int_{0}^{+\infty}k^{D}\hat{u}(k)^{2}dk.
\end{equation}
Equation (\ref{laa2}) also represents the asymptotic behavior of the
diffusion coefficient (\ref{isob9}) for $|{\bf v}|\rightarrow
+\infty$ and for $T\rightarrow +\infty$ since in these limits
$\epsilon\simeq 1$.  In that case, the tensor $G^{\mu\nu}$ given by
Eq. (\ref{laa2}) can be calculated explicitly by introducing
spherical or polar systems of coordinates.  In $D=3$,
\begin{eqnarray}
G^{\mu\nu}=(G_{\|}-{1\over 2}G_{\perp}){x^{\mu}x^{\nu}\over x^{2}}+{1\over 2}G_{\perp}\delta^{\mu\nu},
\label{laa4}
\end{eqnarray}
with
\begin{eqnarray}
G_{\|}={2\pi^{3/2}\over x}G(x),\qquad
G_{\perp}={2\pi^{3/2}\over x}\lbrack {\rm erf}(x)-G(x)\rbrack,
\label{laa5}
\end{eqnarray}
where
\begin{eqnarray}
G(x)={2\over\sqrt{\pi}}{1\over x^{2}}\int_{0}^{x}t^{2}e^{-t^{2}}dt={1\over 2x^{2}}\biggl\lbrack {\rm erf}(x)-{2x\over \sqrt{\pi}}e^{-x^{2}}\biggr\rbrack,
\label{laa6}
\end{eqnarray}
\begin{eqnarray}
{\rm erf}(x)={2\over \sqrt{\pi}}\int_{0}^{x}e^{-t^{2}}dt.
\label{laa7}
\end{eqnarray}
In $D=2$,
\begin{eqnarray}
G^{\mu\nu}=(G_{\|}-G_{\perp}){x^{\mu}x^{\nu}\over x^{2}}+G_{\perp}\delta^{\mu\nu},
\label{laa8}
\end{eqnarray}
with
\begin{eqnarray}
G_{\|}=\pi e^{-{x^{2}\over 2}}\biggl\lbrack I_{0}\biggl ({x^{2}\over 2}\biggr )- I_{1}\biggl ({x^{2}\over 2}\biggr )\biggr\rbrack, \qquad
G_{\perp}=\pi e^{-{x^{2}\over 2}}\biggl\lbrack I_{0}\biggl ({x^{2}\over 2}\biggr )+I_{1}\biggl ({x^{2}\over 2}\biggr )\biggr\rbrack.
\label{laa9}
\end{eqnarray}
In the above expressions $G_{\|}$ and $G_{\perp}$ are the diffusion
coefficients in the directions parallel and perpendicular to the
velocity of the test particle. Finally, in $D=1$,
\begin{eqnarray}
G(x)=2 e^{-x^{2}}.
\label{laa10}
\end{eqnarray}

We can also develop a test particle approach in the case of 2D point
vortices to describes the stochastic evolution of a test vortex in a
bath of field vortices at statistical equilibrium
\cite{drift,kin,houches}. Making a thermal bath approximation, we
can transform the kinetic equation (\ref{landau12}) into a
Fokker-Planck equation of the form
\begin{eqnarray}
{\partial P\over\partial t}={1\over r}{\partial\over\partial r}\biggl\lbrack rD(r)\biggl ({\partial P\over\partial r}+\beta\gamma P{d\psi_{eq}\over dr}\biggr )\biggr\rbrack,
\label{laa11}
\end{eqnarray}
with a diffusion coefficient
\begin{eqnarray}
D(r)={\gamma\over 8}{1\over |\Sigma(r)|}\ln N \langle\omega\rangle_{eq}(r),
\label{laa12}
\end{eqnarray}
where $\Sigma(r)=r\Omega'_{eq}(r)$ is the local shear created by the
statistical distribution of field vortices (the angular velocity
$\Omega_{eq}(r)$ is related to the vorticity by $\omega_{eq}(r)=(1/
r)(r^2\Omega_{eq})'$). In Eq. (\ref{laa11}), $P(r,t)$ denotes the
density probability of finding the test vortex in $r$ at time $t$.
In the more general case where the field vortices have a
distribution $\omega_{0}(r)$ which is a stable stationary solution
of the 2D Euler equation (not necessarily the equilibrium Boltzmann
distribution), the Fokker-Planck equation (\ref{laa11}) is replaced
by
\begin{eqnarray}
{\partial P\over\partial t}={1\over r}{\partial\over\partial
r}\biggl\lbrack rD(r)\biggl ({\partial P\over\partial r}-
P{d \ln \omega_{0}\over d r}\biggr )\biggr\rbrack,
\label{laa13}
\end{eqnarray}
and the diffusion coefficient is still given by Eq. (\ref{laa12})
with $\omega_{0}$ in place of $\langle\omega\rangle_{eq}$. In
particular, for a vorticity profile $\omega_{0}(r)=Ae^{-\lambda
r^2}$ of the field vortices, it is easy to see that the diffusion
coefficient of the test vortex decreases like $D(r)\sim r^2
e^{-\lambda r^2}$ for $r\rightarrow +\infty$. We emphasize the
analogies with the Fokker-Planck equations derived previously in the
case of material particles. Note that for point vortices, the
dynamical friction is replaced by a systematic drift \cite{drift}
along the vorticity gradient.

\subsubsection{The case $D=1$} \label{sec_dun}

In $D=1$, the
Fokker-Planck equation (\ref{tp1}) can be written
\begin{equation}
\label{dun1}  {\partial P\over\partial t}={\partial\over\partial v}\biggl\lbrack D(v)\biggl ({\partial P\over\partial v}-P {d\ln f_{0}\over dv}\biggr )\biggr\rbrack,
\end{equation}
where $D(v)$ is given by
\begin{equation}
\label{dun2} D(v)=4\pi^{2}m f_{0}(v)\int_{0}^{+\infty} dk {k\hat{u}(k)^{2}\over |\epsilon(k,kv)|^{2}}.
\end{equation}
Equation (\ref{dun1}) is similar to a Fokker-Planck equation
describing the motion of a Brownian particle in a potential $U({\bf
v})=-\ln f_{0}({\bf v})$ created by the field particles. In $D=1$,
the distribution function of the test particle $P(v,t)$ always
relaxes toward the distribution of the bath $f_0(v)$ for
$t\rightarrow +\infty$ (the typical timescale governing the approach
of the test particle to the bath distribution is $\sim N t_D$). In
other dimensions, we see from Eq. (\ref{tp1}) that this is true only
when $f_{0}({\bf v})$ is the Maxwellian distribution (statistical
equilibrium state). The particularity of the dimension $D=1$ is that
the Lenard-Balescu collision operator (\ref{bal1}) cancels out for
{\it any} distribution function $f(v)$ while in $D=2$ and $D=3$ the
cancellation of the collision operator occurs only when the
distribution is Maxwellian. In $D=2$ and $D=3$, an arbitrary
distribution of the field particles $f_{0}({\bf v},t)$  changes on a
timescale of order $Nt_{D}$ as it relaxes to the Maxwellian due to
the development of correlations (finite $N$ effects). Therefore, the
test particle approach is valid only for $t\ll Nt_{D}$, in an
interval of times where we can consider that $f_0({\bf v},t)$ is
approximately stationary. But in that case, the distribution of the
test particle $P({\bf v},t)$ does not relax toward $f_0({\bf v})$.
It is only when $f_{0}({\bf v})$ is the Maxwellian that we can
consider that the distribution of the field particles is frozen (for
any time). In that case, $P({\bf v},t)$ relaxes toward $f_{0}({\bf
v})$ in a time $\sim Nt_{D}$. The situation is different in $D=1$.
In $D=1$, a stable stationary solution of the Vlasov equation (any)
does {\it not} change on a timescale of order $N t_{D}$. Therefore,
we can consider that the distribution of the field particles
$f_0(v)$ is frozen on a time $\sim Nt_{D}$ (or larger) which is
precisely the time it takes to a test particle to relax toward
$f_{0}(v)$. The fact that the distribution $P(v,t)$ of a test
particle relaxes to any distribution $f_0(v)$ of the field particles
explains why the distribution of the field particles  does not
change on a time $\sim Nt_{D}$.

For the Maxwellian distribution in $D=1$,
one has
\begin{equation}
\label{dun3} {\partial P\over\partial t}={1\over t_{R}}{\partial\over\partial x}\biggl\lbrack G({x})\biggl ({\partial P\over\partial x}+2 P x\biggr )\biggr\rbrack,
\end{equation}
with
\begin{equation}
\label{dun4} G({x})=2L^{2}\int_{0}^{+\infty} d{k}{k\eta(k)^{2}e^{-{x}^{2}}\over \lbrack 1-\eta(k)B(x)\rbrack^{2}+\pi \eta(k)^{2}{x}^{2}e^{-2{x}^{2}}},
\end{equation}
\begin{equation}
\label{dun5} t_{R}^{-1}={1\over 2}\biggl ({1\over 8\pi}\biggr )^{1/2}{v_{m}\over n}{1\over L^{2}}.
\end{equation}
These results extend those obtained in \cite{bouchet,bd,cvb} for the HMF model for which
\begin{equation}
\label{dun4hmf} G({x})={2L^{2}e^{-{x}^{2}}\over (T/T_{c}-B(x))^{2}+\pi {x}^{2}e^{-2{x}^{2}}}.
\end{equation}
Note that for periodic potentials, the integral over ${\bf k}$ must be replaced by a discrete summation over the different modes.  On the other hand, for a Coulombian potential $\eta(k)=-k_{D}^{2}/k^{2}$, the integration on $k$ can be made explicitly, using $\Phi_{1}(x)=\pi/4-\tan^{-1}(x)/2$ and we obtain
\begin{equation}
\label{gcoul} G({x})={\sqrt{\pi} L^{2} k_{D}^{2}\over 2|x|}\biggl\lbrack 1-{2\over\pi}\tan^{-1}\biggl ({B(x)\over \sqrt{\pi}|x|e^{-x^{2}}}\biggr )\biggr\rbrack,
\end{equation}
with asymptotic behaviors $G(x)\sim \sqrt{\pi}L^{2}k_{D}^{2}/|x|$
for $x\rightarrow \pm\infty$ and $G(0)=L^{2}k_{D}^{2}$.

\subsection{Temporal correlation functions} \label{sec_tc}

We now wish to determine the temporal correlation function of the force
experienced by the test particle. If we ignore collective effects, the
temporal correlation function is given by
\begin{eqnarray}
\label{tc1} \langle F^{\mu}(0)F^{\nu}(t)\rangle=N\int F^{\mu}(1\rightarrow 0,0)F^{\nu}(1\rightarrow 0,t)P_{1}({\bf r}_{1},{\bf v}_{1})d^{D}{\bf r}_{1}d^{D}{\bf v}_{1}.
\end{eqnarray}
Using Fourier transforms and making a linear trajectory approximation as in Appendix \ref{sec_lt},
we obtain
\begin{eqnarray}
\label{tc2} \langle F^{\mu}(0)F^{\nu}(t)\rangle=m (2\pi)^{D}\int k^{\mu}k^{\nu}\hat{u}({\bf k})^{2}e^{-i{\bf k}\cdot {\bf u}t}f_{0}({\bf v}_{1})d^{D}{\bf v}_{1}d^{D}{\bf k}.
\end{eqnarray}
We can also write Eq. (\ref{tc2}) in the form
\begin{eqnarray}
\label{tc3} \langle F^{\mu}(0)F^{\nu}(t)\rangle=m (2\pi)^{2D}\int k^{\mu}k^{\nu}\hat{u}({\bf k})^{2}e^{-i{\bf k}\cdot {\bf v}t}\hat{f}_{0}({\bf k}t)d^{D}{\bf k}.
\end{eqnarray}
For a Maxwellian distribution, we get
\begin{eqnarray}
\label{tc4} \langle F^{\mu}(0)F^{\nu}(t)\rangle=\rho m (2\pi)^{D}\int k^{\mu}k^{\nu}\hat{u}({\bf k})^{2}e^{-i{\bf k}\cdot {\bf v}t}e^{-{k^{2}t^{2}\over 2\beta m}}d^{D}{\bf k}.
\end{eqnarray}
For the gravitational interaction, we can easily perform the
integrations by introducing a spherical system of coordinates. The
correlation function can finally be written
\begin{eqnarray}
C^{\mu\nu}=(C_{\|}-{1\over 2}C_{\perp}){v^{\mu}v^{\nu}\over v^{2}}+{1\over 2}C_{\perp}\delta^{\mu\nu},
\label{tc5}
\end{eqnarray}
with
\begin{eqnarray}
C_{\|}={4\pi \rho m G^{2}\over vt}G(x),
\label{tc6}
\end{eqnarray}
\begin{eqnarray}
C_{\perp}={4\pi \rho m G^{2}\over vt}\lbrack {\rm erf}(x)-G(x)\rbrack.
\label{tc7}
\end{eqnarray}
where ${\bf x}=(\beta m/2)^{1/2}{\bf v}$ and the function $G(x)$ is defined by
Eq. (\ref{laa6}).  We note the result
\begin{eqnarray}
\langle {\bf F}(0)\cdot {\bf F}(t)\rangle={4\pi \rho m G^{2}\over vt}{\rm erf}(x),
\label{tc8}
\end{eqnarray}
which shows that the correlation function of the gravitational force
decreases as $t^{-1}$. This asymptotic result was first noted by
Chandrasekhar \cite{cunsurt} using a different approach. The diffusion
coefficient is then obtained from the Kubo formula (\ref{tp8}) by
integrating over time $t$. This returns Eqs.
(\ref{laa4})-(\ref{laa6}) except that the logarithmic divergence $\int
dk/k$ in Eq. (\ref{laa3}) now appears on the time integration $\int
dt/t$. The divergence at large scales is connected to the very slow
(algebraic) decay of the temporal correlation function. This algebraic
decay is strikingly in contrast with usual Markov processes in which
the correlations decrease exponentially rapidly with time. Coming back
to Eq. (\ref{tc4}), we note that each mode has a {gaussian} decay
$\sim {\rm exp}\lbrack -{k^{2}t^{2}/ 2\beta m}\rbrack$ but the
integration over all modes leads to an algebraic behavior $\sim
t^{-1}$.

If we come back to the general problem and take into account collective effects, it is shown in Appendix \ref{sec_leba} that
\begin{eqnarray}
\label{tc9} \langle F^{\mu}(0)F^{\nu}(t)\rangle=m (2\pi)^{D}\int k^{\mu}k^{\nu}{\hat{u}({\bf k})^{2}\over |\epsilon({\bf k},{\bf k}\cdot {\bf v}_{1})|^{2}}e^{-i{\bf k}\cdot {\bf u}t}f_{0}({\bf v}_{1})d^{D}{\bf v}_{1}d^{D}{\bf k}.
\end{eqnarray}
If we integrate over time $t$ and use the Kubo formula (\ref{tp8}),  we recover the expression (\ref{tp4}) of the diffusion coefficient. Introducing
\begin{eqnarray}
\label{tc10} Q({\bf k},t)=\int e^{i{\bf k}\cdot {\bf v}_{1}t}{f_{0}({\bf v}_{1})\over |\epsilon({\bf k},{\bf k}\cdot {\bf v}_{1})|^{2}}d^{D}{\bf v}_{1},
\end{eqnarray}
we can write the correlation function more compactly as
\begin{eqnarray}
\label{tc11} \langle F^{\mu}(0)F^{\nu}(t)\rangle=m (2\pi)^{D}\int k^{\mu}k^{\nu}{\hat{u}({\bf k})^{2}}e^{-i{\bf k}\cdot {\bf v}t}Q({\bf k},t)d^{D}{\bf k}.
\end{eqnarray}
For a Maxwellian distribution (\ref{isob1}),
\begin{eqnarray}
\label{tc12} Q({\bf k},t)=\rho \biggl ({\beta m\over 2\pi}\biggr )^{1/2}{1\over k}\int_{-\infty}^{+\infty} e^{i\omega t}{e^{-\beta m {\omega^{2}\over 2k^{2}}}\over |\epsilon(k,\omega)|^{2}}d\omega,
\end{eqnarray}
where we have chosen ${\bf k}$ in the direction of $x$ and set
$\omega=kv_{x}$.
To determine the large time asymptotics of Eq. (\ref{tc12}), we need to investigate the poles of the function
\begin{eqnarray}
\label{tc13} f(\omega)={e^{-\beta m {\omega^{2}\over 2k^{2}}}\over |\epsilon|^{2}(k,\omega)}.
\end{eqnarray}
Using the notations introduced in  Eq. (\ref{isob9}), we have
\begin{eqnarray}
\label{tc14} |\epsilon|^{2}(k,\omega)=\biggl\lbrack 1-\eta(k)B\biggl (\sqrt{\beta m\over 2}{\omega\over k}\biggr )\biggr\rbrack^{2}+\eta(k)^{2}{\pi\over 2} {\beta m}\biggl ({\omega\over k}\biggr )^{2}e^{-{\beta m\omega^{2}\over 2k^{2}}}.
\end{eqnarray}
Setting $\omega=i\lambda$ where $\lambda$ is real, we find after
straightforward algebra that the foregoing expression can be rewritten
\begin{eqnarray}
\label{tc15} |\epsilon|^{2}(k,i\lambda)=\epsilon(k,i\lambda)\epsilon(k,-i\lambda),
\end{eqnarray}
where we recall that
\begin{eqnarray}
\label{tc16}\epsilon(k,i\lambda)=1-\eta(k)/G\biggl (\sqrt{\beta m\over 2}{\lambda\over k}\biggr ).
\end{eqnarray}
Equation (\ref{tc13}) can thus be rewritten
\begin{eqnarray}
\label{tc17} f(\omega)={e^{\beta m {\lambda^{2}\over 2k^{2}}}\over  \epsilon(k,i\lambda)\epsilon(k,-i\lambda)}.
\end{eqnarray}
Clearly, this is an even function of $\lambda$. We need to determine
the values of $\lambda$ for which the denominator vanishes. Since the
distribution of the bath is stable by hypothesis ($T>T_{c}$), we have $\eta(k)<1$ for all $k$. Therefore, the denominator
cancels out for $\lambda=\pm \gamma$ where $\gamma>0$ is determined by
$\epsilon(k,-i\gamma)=0$. Thus $\gamma$ coincides with the damping
rate of the stable perturbed solutions of the Vlasov equation (see
Sec. \ref{sec_ls}). It is given as a function of $k$ by
Eq. (\ref{ls9}). Next, we consider $\lambda=\pm\gamma+\Delta$ where
$\Delta\ll 1$. Expanding Eq. (\ref{tc17}) for small values of
$\Delta$, we find after elementary calculations that for $\omega\rightarrow \pm i\gamma$
\begin{equation}
\label{tc20} f(\omega)\sim {K(\gamma)\over \omega^{2}+\gamma^{2}},
\end{equation}
where the constant is given by
\begin{equation}
\label{tc20bis} K(\gamma)={2\over\sqrt{\pi}}{k^{2}\over\beta m}{1\over |F' (\sqrt{\beta m\over 2}{\gamma\over k} )|}.
\end{equation}
Since $f(\omega)$ behaves like a Lorentzian for $\omega\rightarrow \pm i\gamma$, this implies that, for $t\rightarrow +\infty$,  the different modes in the correlation function decrease
exponentially rapidly as
\begin{equation}
\label{tc21} Q({\bf k},t)\sim \rho\sqrt{2\over\beta m}{k\over\gamma_{k}}{1\over |F' (\sqrt{\beta m\over 2}{\gamma_{k}\over k} )|} e^{-\gamma_{k}t},
\end{equation}
with a rate
\begin{equation}
\label{tc22}\gamma_{k}=\sqrt{2\over\beta m}k F^{-1}\lbrack \eta(k)\rbrack
\end{equation}
depending on the wave vector $k$ (see Eq. (\ref{ls9})). These results
generalize those obtained in \cite{bouchet,cvb} for the HMF model. In this case, the correlation
function decreases like
\begin{eqnarray}
\label{tc25} \langle F(0)F(t)\rangle\sim {k^{2}M\over 8\pi^{2}}{\sqrt{2T}\over\gamma}{1\over |F'({\gamma\over \sqrt{2T}})|}\cos(vt)e^{-\gamma t},
\end{eqnarray}
where the decay rate
\begin{eqnarray}
\label{tc26}\gamma=({2/ \beta})^{1/2} F^{-1}(\eta)
\end{eqnarray}
only depends on the temperature (recall that $\eta=\beta
kM/4\pi$). For $T\rightarrow T_{c}^{+}$,
$\gamma\sim (8/kM)^{1/2}(T-T_{c})$
and for $T\rightarrow +\infty$, $\gamma\sim \sqrt{2T\ln T}$
(to leading order). Close to the critical temperature,
the correlation function decreases very slowly. {\it Note that this slow
decay may invalidate the Markovian approximation close to the critical
point and lead to dynamical anomalies.} On the other hand, at high
temperatures, the decay is very fast. In fact, had we ignored
collective effects and used Eq. (\ref{tc4}), we would have obtained a
gaussian decay
\begin{equation}
\label{tc27}\langle F(0) F(t)\rangle={\rho k^{2}\over 4\pi}\cos(vt)e^{-{t^{2}\over 2\beta}},
\end{equation}
instead of an exponential decay with large $\gamma$-rate. This shows
that collective effects are important even far from the critical
point since they modify the large time behavior of the temporal
correlation function. For the gravitational potential, Eq.
(\ref{tc22}) becomes
\begin{equation}
\label{tc28}\gamma_{k}=\sqrt{2\over\beta m} k F^{-1}\biggl ( {k_{J}^{2}\over k^{2}}\biggr ).
\end{equation}
Using $F^{-1}(x)={1\over\sqrt{\pi}}(1-x)+...$ for $x\rightarrow 1^{-}$ and $F^{-1}(x)\sim \sqrt{-\ln x}$ for $x\rightarrow 0$, we find that
\begin{equation}
\label{tc29}\gamma_{k}\simeq \sqrt{2\over\pi\beta m} k_{J}\biggl (1- {k_{J}^{2}\over k^{2}}\biggr ), \qquad (k\rightarrow k_{J}^{+})
\end{equation}
\begin{equation}
\label{tc30}\gamma_{k}\sim {2k\over\sqrt{\beta m}}\ln\biggl ({k\over k_{J}}\biggr )^{1/2},\qquad  (k\rightarrow +\infty).
\end{equation}

\subsection{Evolution of the spatial correlation function in the linear regime} \label{sec_ge}

To conclude this section on the kinetic theory of Hamiltonian systems
with long-range interactions, we would like to discuss the time
evolution of the spatial correlation function. Returning to the second
equation of the BBGKY hierarchy (\ref{h5}) for the two-body
correlation function and considering the case of a homogeneous medium,
we obtain
\begin{eqnarray}
\label{ev1}  {\partial P_{2}'\over\partial t}+{\bf v}_{1}{\partial P_{2}'\over\partial {\bf r}_{1}}+{\bf F}(2\rightarrow 1)P_{1}({\bf v}_{2}){\partial P_{1}\over\partial {\bf v}_{1}}({\bf v}_{1})\nonumber\\
+N{\partial\over\partial {\bf v}_{1}}\int {\bf F}(3\rightarrow 1)P_{2}'({\bf x}_{2},{\bf x}_{3},t)P_{1}({\bf v}_{1})d^{D}{\bf x}_{3}+(1\leftrightarrow 2)=0,
\end{eqnarray}
while the first equation (\ref{h4}) gives $\partial P_{1}/\partial t=0$ for $N\rightarrow +\infty$. We assume that, initially, no correlation is
present among the particles. Then, for sufficiently small times
(linear regime), the correlations will be small and we can neglect the
integrals in Eq. (\ref{ev1}). This yields
\begin{equation}
\label{ev2} {\partial P_{2}'\over\partial t}+{\bf v}_{1}{\partial P_{2}'\over\partial {\bf r}_{1}}+{\bf F}(2\rightarrow 1)P_{1}({\bf v}_{2}){\partial P_{1}\over\partial {\bf v}_{1}}({\bf v}_{1})+(1\leftrightarrow 2)=0.
\end{equation}
We now assume that, initially, $P_1\sim {\rm exp}\lbrack -\beta m v^{2}/
2\rbrack$ is the Maxwellian distribution. As we have seen previously, this
distribution is conserved to leading order at later times.
Introducing the correlation function $h$ though the defining relation
\begin{equation}
\label{ev3} P_{2}({\bf r}_{1},{\bf v}_{1},{\bf r}_{2},{\bf v}_{2},t)=P_{1}({\bf v}_{1})P_{1}({\bf v}_{2})\lbrack 1+h({\bf r}_{1}-{\bf r}_{2},{\bf v}_{1}-{\bf v}_{2},t)\rbrack,
\end{equation}
we find that it satisfies the
differential equation
\begin{equation}
\label{ev4} {\partial h\over\partial t}+{\bf u}\cdot {\partial h\over\partial {\bf x}}=\beta m {\bf F}(2\rightarrow 1)\cdot {\bf u},
\end{equation}
where ${\bf x}={\bf r}_{1}-{\bf r}_{2}$ and ${\bf u}={\bf v}_{1}-{\bf
v}_{2}$. For a stationary solution, we recover Eq. (I-53) of Paper
I. Note also that Eq. (I-51) can be obtained from the static
($\partial/\partial t=0$) expression of Eq. (\ref{ev1}) by keeping the
integrals and assuming that $P_{1}$ is Maxwellian.

Returning to Eq. (\ref{ev4}), taking its Fourier transform and solving
the resulting first order differential equation, we get
\begin{equation}
\label{ev5}\hat{h}({\bf k},{\bf u},t)=-\beta m^{2}\hat{u}(k)(1-e^{-i{\bf k}\cdot {\bf u}t}),
\end{equation}
where we have assumed that, initially, the system is uncorrelated.
Taking the inverse Fourier transform of Eq. (\ref{ev5}), we finally
obtain
\begin{equation}
\label{ev6}{h}({\bf x},{\bf u},t)=\beta m^{2}\lbrack u({\bf x}-{\bf u}t)-u({\bf x})\rbrack.
\end{equation}
It may be useful to discuss the  HMF model explicitly. In that case,
Eq. (\ref{ev4}) becomes
\begin{equation}
\label{ev7} {\partial h\over\partial t}+u{\partial h\over\partial \phi}=-{k\beta \over 2\pi}u\sin\phi.
\end{equation}
The stationary solution of Eq. (\ref{ev7}) is
\begin{equation}
\label{ev8}h(\phi)={k\beta\over 2\pi}\cos\phi.
\end{equation}
This is similar to Eq. (I-66) of Paper I, but it does not display the
expected divergence as we approach the critical point $T_{c}$. This is
due to the neglect of the integrals in Eq. (\ref{ev1}).  Furthermore,
it is noteworthy that the solution of Eq. (\ref{ev7}) does not converge
towards the stationary distribution (\ref{ev8}) for $t\rightarrow +\infty$. Indeed, the explicit time-dependent solution of Eq. (\ref{ev7}) is
\begin{equation}
\label{ev9} h(\phi,u,t)={k\beta\over 2\pi}\biggl\lbrack \cos\phi-\cos(\phi-ut)\biggr\rbrack,
\end{equation}
which shows an oscillatory behavior. This is not really surprising
since Eq. (\ref{ev9}) is only valid for short times. In this linear
regime
\begin{equation}
\label{ev10} h(\phi,u,t)=-{k\beta\over 2\pi}\sin(\phi) u t \qquad (t\rightarrow 0).
\end{equation}

\section{Kinetic theory of Brownian systems}
\label{sec_gt}

\subsection{The non-local Kramers equation}
\label{sec_k}

We shall now derive the kinetic equations describing a system of
Brownian particles with long-range interactions defined by the
stochastic equations (I-39)-(I-40) of Paper I. We shall use and generalize
the method of Martzel \& Aslangul \cite{ma,mar}. We start from the
general Markov process
\begin{eqnarray}
\label{k1}
P_{N}({\bf r}_{1},{\bf v}_{1},...,{\bf r}_{N},{\bf v}_{N},t+\Delta t)=\int d^{D}(\Delta {\bf r}_{1})d^{D}(\Delta {\bf v}_{1})...d^{D}(\Delta {\bf r}_{N})d^{D}(\Delta {\bf v}_{N})\nonumber\\
{\times} P_{N}({\bf r}_{1}-\Delta {\bf r}_{1},{\bf v}_{1}-\Delta {\bf
v}_{1},...,{\bf r}_{N}
-\Delta {\bf r}_{N},{\bf v}_{N}-\Delta {\bf v}_{N},t)\nonumber\\
{\times} w({\bf r}_{1}-\Delta {\bf r}_{1},{\bf v}_{1}-\Delta {\bf
v}_{1},...,{\bf r}_{N}-\Delta {\bf r}_{N},{\bf v}_{N}-\Delta {\bf
v}_{N} |\Delta {\bf r}_{1},\Delta {\bf v}_{1},...,\Delta {\bf
r}_{N},\Delta {\bf v}_{N}    ).
\end{eqnarray}
where $w$ denotes the transition probability from one state to the
other specified by the term in parenthesis. Using Eq. (I-39), it can be rewritten
\begin{eqnarray}
\label{k2}
 w({\bf r}_{1},{\bf v}_{1},...,{\bf
r}_{N},{\bf
v}_{N} |\Delta {\bf r}_{1},\Delta {\bf v}_{1},...,\Delta {\bf r}_{N},\Delta {\bf v}_{N})=\nonumber\\
\delta(\Delta {\bf r}_{1}-{\bf v}_{1}\Delta t)...\delta(\Delta
{\bf r}_{N}-{\bf v}_{N}\Delta t)\psi({\bf r}_{1},{\bf
v}_{1},...,{\bf r}_{N},{\bf v}_{N} |\Delta {\bf v}_{1},...,\Delta
{\bf v}_{N}).
\end{eqnarray}
Then, the integration over $\Delta{\bf r}_{1}$...$\Delta{\bf r}_{N}$ is
straightforward and yields
\begin{eqnarray}
\label{k3}
P_{N}({\bf r}_{1},{\bf v}_{1},...,{\bf r}_{N},{\bf v}_{N},t+\Delta t)=\int d^{D}(\Delta {\bf v}_{1})...d^{D}(\Delta {\bf v}_{N})\nonumber\\
{\times} P_{N}({\bf r}_{1}-{\bf v}_{1}\Delta t,{\bf v}_{1}-\Delta {\bf
v}_{1},...,{\bf r}_{N}
-{\bf v}_{N}\Delta t,{\bf v}_{N}-\Delta {\bf v}_{N},t)\nonumber\\
{\times} \psi({\bf r}_{1}-{\bf v}_{1}\Delta t,{\bf v}_{1}-\Delta {\bf
v}_{1},...,{\bf r}_{N}-{\bf v}_{N}\Delta t,{\bf v}_{N}-\Delta {\bf
v}_{N} |\Delta {\bf v}_{1},...,\Delta {\bf v}_{N}    ),
\end{eqnarray}
or equivalently
\begin{eqnarray}
\label{k4}
P_{N}({\bf r}_{1}+{\bf v}_{1}\Delta t,{\bf v}_{1},...,{\bf r}_{N}+{\bf v}_{N}\Delta t,{\bf v}_{N},t+\Delta t)=\int d^{D}(\Delta {\bf v}_{1})...d^{D}(\Delta {\bf v}_{N})\nonumber\\
{\times} P_{N}({\bf
r}_{1},{\bf v}_{1}-\Delta {\bf v}_{1},...,{\bf r}_{N},{\bf v}_{N}-\Delta {\bf v}_{N},t)\nonumber\\
{\times} \psi({\bf r}_{1},{\bf v}_{1}-\Delta {\bf v}_{1},...,{\bf
r}_{N},{\bf v}_{N}-\Delta {\bf v}_{N} |\Delta {\bf
v}_{1},...,\Delta {\bf v}_{N}    ).
\end{eqnarray}
Expanding the right hand side in Taylor series and introducing the
Kramers-Moyal moments
\begin{eqnarray}
\label{k5} M_{n_{1}...n_{N}}({\bf r}_{1},{\bf v}_{1},...,{\bf
r}_{N},{\bf v}_{N})={\rm lim}_{\Delta t\rightarrow 0}{1\over
\Delta t\ n_{1}!...n_{N}!}\int d^{D}(\Delta {\bf
v}_{1})...d^{D}(\Delta {\bf v}_{N})\nonumber\\
 {\times} (-\Delta {\bf
v}_{1})^{n_{1}}...(-\Delta {\bf v}_{N})^{n_{N}}
 \psi({\bf r}_{1},{\bf v}_{1},...,{\bf r}_{N},{\bf v}_{N}|\Delta {\bf v}_{1},...,\Delta {\bf
v}_{N}),
\end{eqnarray}
we get
\begin{equation}
\label{k6} {\partial P_{N}\over\partial t}+\sum_{i=1}^{N}{\bf
v}_{i}{\partial P_{N}\over\partial {\bf
r}_{i}}=\sum_{n_{1}...n_{N}}{\partial^{n_{1}}\over\partial {\bf
v}_{1}^{n_{1}}}...{\partial^{n_{N}}\over\partial {\bf
v}_{N}^{n_{N}}}\lbrace M_{n_{1}...n_{N}}P_{N}\rbrace,
\end{equation}
where the sum runs over all indices such that $\sum_{i}n_{i}\ge
1$. For the stochastic process (I-39)-(I-40), only a
few moments do not vanish, namely
\begin{equation}
\label{k7} M_{0...n_{i}=1...0}=\xi {\bf v}_{i}+m\nabla_{i}U({\bf
r}_{1},...,{\bf r}_{N}),
\end{equation}
\begin{equation}
\label{k8} M_{0...n_{i}=2...0}=D.
\end{equation}
Substituting these results in Eq. (\ref{k6}), we obtain the $N$-body Fokker-Planck equation
\begin{equation}
\label{k9} {\partial P_{N}\over\partial t}+\sum_{i=1}^{N}\biggl
({\bf v}_{i}{\partial P_{N}\over\partial {\bf r}_{i}}+{\bf
F}_{i}{\partial P_{N}\over\partial {\bf v}_{i}}\biggr
)=\sum_{i=1}^{N} {\partial\over\partial {\bf v}_{i}}\biggl\lbrack
D{\partial P_{N}\over\partial {\bf v}_{i}}+\xi P_{N}{\bf
v}_{i}\biggr\rbrack,
\end{equation}
where ${\bf F}_{i}=-m\nabla_{i}U({\bf r}_{1},...,{\bf r}_{N})$ is the
force by unit of mass acting on the $i$-th particle. The $N$-body Fokker-Planck equation decreases the free energy
\begin{equation}
\label{k9alpha}F[P_{N}]=\langle E\rangle[P_{N}]-TS[P_{N}]
\end{equation}
constructed with the average energy (I-8) and the entropy (I-26) defined
in Paper I. Indeed, an explicit calculation yields
\begin{equation}
\label{k9beta}\dot F=-\sum_{i=1}^{N}\int {1\over\xi P_{N}}\biggl
(D{\partial P_{N}\over\partial {\bf v}_{i}}+\xi P_{N}{\bf
v}_{i}\biggr )^{2}d^{D}{\bf r}_{1}d^{D}{\bf v}_{1}...d^{D}{\bf r}_{N}d^{D}{\bf v}_{N}\le 0.
\end{equation}
Since $\dot F=0$ at equilibrium, the term in bracket in Eq. (\ref{k9})
vanishes by virtue of Eq. (\ref{k9beta}). Since $\partial/\partial
t=0$, the advective term must also vanish, independently. From these
two requirements, we find that the stationary solution of the $N$-body
Fokker-Planck equation corresponds to the canonical distribution
(I-42)  which minimizes the free energy.

We can now obtain the
equivalent of the BBGKY hierarchy for the reduced distribution
functions. It reads
\begin{eqnarray}
\label{k10} {\partial P_{j}\over\partial t}+\sum_{i=1}^{j}{\bf v}_{i}{\partial
P_{j}\over\partial {\bf r}_{i}}+\sum_{i=1}^{j}\sum_{k=1,k\neq i}^{j} {\bf F}(k\rightarrow i){\partial P_{j}\over \partial {\bf v}_{i}}+(N-j)\sum_{i=1}^{j}\int d^{D}{\bf x}_{j+1}{\bf F}(j+1\rightarrow i){\partial P_{j+1}\over\partial {\bf v}_{i}}\nonumber\\
=\sum_{i=1}^{j}{\partial\over\partial {\bf v}_{i}}\biggl\lbrack D{\partial P_{j}\over\partial {\bf v}_{i}}+\xi P_{j}{\bf v}_{i}\biggr\rbrack.
\end{eqnarray}
In particular, the first equation of the hierarchy is
\begin{eqnarray}
\label{k11} {\partial P_{1}\over\partial t}+{\bf v}_{1}{\partial
P_{1}\over\partial {\bf r}_{1}}+(N-1)\int d^{D}{\bf x}_{2}{\bf F}(2\rightarrow 1){\partial P_{2}\over\partial {\bf v}_{1}}
={\partial\over\partial {\bf v}_{1}}\biggl\lbrack D{\partial P_{1}\over\partial {\bf v}_{1}}+\xi P_{1}{\bf v}_{1}\biggr\rbrack.
\end{eqnarray}
We can similarly obtain an equation for the two-body correlation
function.  Implementing the decomposition (I-14)-(I-15) of Paper I,
neglecting the cumulant of the three-body correlation function, and taking the
thermodynamical limit defined in Paper I, we find that $P_2'\sim
1/N$. Therefore, in the limit $N\rightarrow +\infty$, we can make the
mean-field approximation
\begin{equation}
\label{k12} P_{2}({\bf r}_{1},{\bf v}_{1},{\bf r}_{2},{\bf
v}_{2},t)=P_{1}({\bf r}_{1},{\bf v}_{1},t)P_{1}({\bf r}_{2},{\bf
v}_{2},t).
\end{equation}
We thus obtain
\begin{equation}
\label{k13} {\partial P_{1}\over\partial t}+{\bf v}_{1}{\partial
P_{1}\over\partial {\bf r}_{1}}+\langle {\bf
F}\rangle_{1}{\partial P_{1}\over\partial {\bf v}_{1}}=
{\partial\over\partial {\bf v}_{1}}\biggl\lbrack
D {\partial P_{1}\over\partial {\bf v}_{1}}+\xi P_{1}{\bf
v}_{1}\biggr\rbrack,
\end{equation}
where
\begin{equation}
\label{k14} \langle {\bf F}\rangle_{1}=-Nm\int d^{D}{\bf
r}_{2}d^{D}{\bf v}_{2} \ {\partial u\over\partial {\bf
r}_{1}}({\bf r}_{1}-{\bf r}_{2})P_{1}({\bf r}_{2},{\bf v}_{2},t).
\end{equation}
Introducing the distribution function $f=NmP_{1}$, this can be rewritten
\begin{equation}
\label{k15} {\partial f\over\partial t}+{\bf v}\cdot {\partial
f\over\partial {\bf r}}+\langle{\bf F}\rangle \cdot {\partial
f\over\partial {\bf v}}={\partial\over\partial {\bf v}}\cdot \biggl
\lbrack D\biggl ({\partial f\over\partial {\bf v}}+\beta m
f{\bf v}\biggr )\biggr\rbrack,
\end{equation}
where
\begin{equation}
\label{k16} \langle{\bf F}\rangle=-\nabla\Phi=-\int d^{D}{\bf r}'
\ \rho({\bf r}',t){\partial u\over\partial {\bf r}}({\bf r}-{\bf
r}')
\end{equation}
is the mean-field force acting on a particle and we have set
$\xi=D\beta m$. Equation (\ref{k15}) is a non-local Kramers equation. It
decreases the free energy
\begin{equation}
\label{k17} F[f]=E[f]-TS[f]={1\over 2}\int f v^{2} d^{D}{\bf r} d^{D}{\bf
v}+ {1\over 2}\int \rho\Phi d^{D}{\bf r}+T\int f\ln f d^{D}{\bf r}
d^{D}{\bf v},
\end{equation}
which plays the role of a Lyapunov functional \cite{gt}. The free
energy (\ref{k17}) is the resulting expression of Eq. (\ref{k9alpha})
in the mean-field approximation. We note that the non-local Kramers
equation (\ref{k15}) is obtained at the same level of approximation
(i.e. for $N\rightarrow +\infty$) as the Vlasov equation
(\ref{vlasov1}) for Hamiltonian systems. The introduction of a
friction and a random force in the equations of motion
(I-39)-(I-40) yields a ``collision term" of the
Fokker-Planck form in the right hand side of Eq. (\ref{k15}). This
``collision'' term selects the mean-field Maxwell-Boltzmann equilibrium
distribution (I-24) among the infinite class of stationary
solutions of the Vlasov equation (left hand side). This mean-field
Maxwell-Boltzmann distribution extremizes the free energy (\ref{k17})
at fixed mass. Furthermore, only minima of free energy are linearly
dynamically stable via Eq. (\ref{k15}) \cite{gt}. We should also
contrast the non-local Fokker-Planck equation (\ref{k15}) to the local
Fokker-Planck Eq. (\ref{isob2}). Although they look similar, their
physical content is quite different. Indeed, Eq. (\ref{tp3}) describes
the motion of a {\it single} test particle in a thermal bath at
statistical equilibrium while Eq. (\ref{k15}) describes the evolution
of the {\it whole} system of $N$ Brownian particles in interaction out
of equilibrium. Therefore, Eq. (\ref{tp3}) is a local differential
equation while Eq. (\ref{k15}) is non-local due to the mean-field
force produced by the distribution function $f({\bf r},{\bf v},t)$
evolving in time. In addition, in the present case, the diffusion
coefficient $D$ is {\it given} in Eq. (I-40) while in the test
particle approach it is derived from the Hamiltonian dynamics at the
order $1/N$ for $N\gg 1$.

\subsection{The non-local Smoluchowski equation}
\label{sec_s}

In the strong friction limit $\xi\rightarrow +\infty$, or equivalently
for large times $t\gg \xi^{-1}$, it is possible to neglect the
inertial term in Eq. (I-40). In that case, we are led to consider a
system of $N$ Brownian particles in interaction described by the
coupled stochastic equations in physical space
\begin{equation}
\label{s1} {d{\bf r}_{i}\over dt}=-\mu m^{2}\nabla_{i}U({\bf
r}_{1},...,{\bf r}_{N})+\sqrt{2D_{*}}{\bf R}_{i}(t),
\end{equation}
where $\mu=1/m\xi$ is the mobility and $D_{*}=D/\xi^{2}=T/m\xi$ is the
diffusion coefficient in physical space. The Einstein relation reads
$\mu=\beta D_{*}$. We can now repeat the above procedure to obtain the
kinetic equations governing the evolution of the Brownian system in
the overdamped regime.  For a general Markovian process in physical
space, we have
\begin{eqnarray}
\label{s2} P_{N}({\bf r}_{1},...,{\bf r}_{N},t+\Delta t)=\int
d^{D}(\Delta {\bf r}_{1})...d^{D}(\Delta {\bf r}_{N})\ P_{N}({\bf
r}_{1}-\Delta {\bf r}_{1},...,{\bf r}_{N}-\Delta {\bf
r}_{N},t)\nonumber\\
{\times} w({\bf r}_{1}-\Delta {\bf r}_{1},...,{\bf r}_{N}-\Delta {\bf
r}_{N}|\Delta {\bf r}_{1},...,\Delta {\bf r}_{N})
\end{eqnarray}
where $w$ denotes the transition probability. Expanding
the right hand side in Taylor series and introducing the
Kramers-Moyal moments
\begin{eqnarray}
\label{s3} M_{n_{1}...n_{N}}({\bf r}_{1},...,{\bf r}_{N})={\rm
lim}_{\Delta t\rightarrow 0}{1\over \Delta t\ n_{1}!...n_{N}!}\int
d^{D}(\Delta {\bf r}_{1})...d^{D}(\Delta {\bf r}_{N})\ (-\Delta
{\bf
r}_{1})^{n_{1}}...(-\Delta {\bf r}_{N})^{n_{N}} \nonumber\\
{\times} w({\bf r}_{1},...,{\bf r}_{N}|\Delta {\bf r}_{1},...,\Delta {\bf
r}_{N}),
\end{eqnarray}
we get
\begin{equation}
\label{s4} {\partial P_{N}\over\partial
t}=\sum_{n_{1}...n_{N}}{\partial^{n_{1}}\over\partial {\bf
r}_{1}^{n_{1}}}...{\partial^{n_{N}}\over\partial {\bf
r}_{N}^{n_{N}}}\lbrace M_{n_{1}...n_{N}}P_{N}\rbrace
\end{equation}
where the sum runs over all indices such that $\sum_{i}n_{i}\ge
1$. For the stochastic process (\ref{s1}), only a few moments do not
vanish, namely
\begin{equation}
\label{s5} M_{0...n_{i}=1...0}=\mu m^{2}\nabla_{i}U({\bf
r}_{1},...,{\bf r}_{N}),
\end{equation}
\begin{equation}
\label{s6} M_{0...n_{i}=2...0}=D_{*}.
\end{equation}
Substituting these results in Eq.
(\ref{s4}), we obtain the $N$-body Fokker-Planck equation
\begin{equation}
\label{s7} {\partial P_{N}\over\partial t}=\sum_{i=1}^{N}
{\partial\over\partial {\bf r}_{i}}\biggl\lbrack
D_{*}{\partial P_{N}\over\partial {\bf r}_{i}}+\mu m^{2}
P_{N}{\partial\over\partial {\bf r}_{i}}U({\bf r}_{1},...,{\bf
r}_{N})\biggr\rbrack.
\end{equation}
The stationary solutions of this equation correspond to the
configurational part of the canonical distribution
(I-44). Again, it is possible to derive the equivalent of the
BBGKY hierarchy for the reduced distribution functions. It reads
\begin{equation}
\label{s8} {\partial P_{j}\over\partial t}=\sum_{i=1}^{j}
{\partial\over\partial {\bf r}_{i}}\biggl\lbrack D_{*}{\partial
P_{j}\over\partial {\bf r}_{i}}+\mu m^{2}\sum_{k=1,k\neq i}^{j} P_{j}{\partial u_{ik}\over\partial {\bf r}_{i}}+\mu m^{2}(N-j)\int P_{j+1} {\partial
u_{i,j+1}\over\partial {\bf r}_{i}}d^{D}{\bf r}_{j+1}\biggr\rbrack.
\end{equation}
Introducing the decomposition (I-14)-(I-15), neglecting the three-body
correlation function and considering only terms of order $1/N$ or
larger in the thermodynamic limit $N\rightarrow +\infty$, we obtain
\begin{equation}
\label{s9}  {\partial P_{1}\over\partial t}=
{\partial\over\partial {\bf r}_{1}}\biggl\lbrack D_{*}{\partial
P_{1}\over\partial {\bf r}_{1}}+\mu Nm^{2} P_{1}({\bf r}_{1},t)\int
P_{1}({\bf r}_{2},t) {\partial u_{12}\over\partial {\bf r}_{1}}
d^{D}{\bf r}_{2}+\mu Nm^{2}\int P'_{2}({\bf r}_{1},{\bf r}_{2},t){\partial
u_{12}\over\partial {\bf r}_{1}}d^{D}{\bf r}_{2}\biggr\rbrack,
\end{equation}
\begin{eqnarray}
\label{s10}{\partial P_{2}'\over\partial t}={\partial\over\partial {\bf r}_{1}}\biggl\lbrack D_{*}{\partial P_{2}'\over\partial {\bf r}_{1}}+\mu m^{2} P_{1}({\bf r}_{1},t) P_{1}({\bf r}_{2},t){\partial u_{12}\over\partial {\bf r}_{1}}+N\mu m^{2} P_{2}'({\bf r}_{1},{\bf r}_{2},t)\int P_{1}({\bf r}_{3},t){\partial u_{13}\over\partial {\bf r}_{1}}d^{D}{\bf r}_{3}\nonumber\\
+N\mu m^{2} P_{1}({\bf r}_{1},t)\int P_{2}'({\bf r}_{2},{\bf r}_{3},t){\partial u_{13}\over\partial {\bf r}_{1}}d^{D}{\bf r}_{3}\biggr\rbrack+(1\leftrightarrow 2).
\end{eqnarray}
The stationary solutions of these equations coincide with the
equations of the equilibrium BBGKY-like hierarchy (see Paper I) as
expected. In the limit $N\rightarrow \infty$, we can make the
mean-field approximation
\begin{equation}
\label{s11} P_{2}({\bf r}_{1},{\bf r}_{2},t)=P_{1}({\bf
r}_{1},t)P_{1}({\bf r}_{2},t).
\end{equation}
The equation for the density then  reduces to
\begin{equation}
\label{s12} {\partial P_{1}\over\partial t}=
{\partial\over\partial {\bf r}_{1}}\biggl\lbrack D_{*}
{\partial P_{1}\over\partial {\bf r}_{1}}+\mu m P_{1}
{\partial\Phi\over\partial {\bf r}_{1}}\biggr\rbrack,
\end{equation}
which can be written
\begin{equation}
\label{s13} {\partial \rho\over\partial t}= \nabla\cdot \lbrack D_{*} (\nabla\rho+\beta m \rho \nabla\Phi)\rbrack,
\end{equation}
where $\Phi({\bf r},t)$ is related to $\rho({\bf r},t)$ as in
Eq. (\ref{k16}).  This is the non-local Smoluchowski equation. This
equation decreases the free energy
\begin{equation}
\label{s14} F[\rho]={1\over 2}\int
\rho\Phi d^{D}{\bf r}+T\int \rho\ln\rho d^{D}{\bf r},
\end{equation}
which plays the role of a Lyapunov functional. The Smoluchowski
equation can also be derived directly from the Kramers equation in the
strong friction limit $\xi\rightarrow +\infty$ \cite{risken}. This can be done by working out the moments equations of Eq. (\ref{k15}) as in
\cite{gt} or by using a Chapman-Enskog expansion as in \cite{lemou}.
To leading order in $1/\xi$, the velocity distribution is Maxwellian
\begin{equation} \label{s14b}
f({\bf r},{\bf v},t)=\biggl ({\beta m\over 2\pi}\biggr )^{D/2}\rho({\bf r},t) e^{-\beta m {v^{2}\over 2}}+O(\xi^{-1}),
\end{equation}
and the evolution of $\rho({\bf r},t)$ is given by
Eq. (\ref{s13}). The free energy (\ref{s14}) can be obtained from
Eq. (\ref{k17}) by using the approximate expression (\ref{s14b}) of
the distribution function to express the free energy as a functional
of $\rho$.

Considering the linear dynamical stability of a spatially homogeneous
distribution of particles with respect to the non-local Smoluchowski
equation (\ref{s13}) with $\Phi=\rho * u$, we immediately get the
dispersion relation
\begin{equation}
\label{s15}
i\xi\omega={T\over m}k^{2}+(2\pi)^{D}\hat{u}(k)k^{2}\rho.
\end{equation}
We note that the condition of instability (corresponding to
$i\omega<0$) is the same as for an isothermal distribution described
by the Euler or the Vlasov equation (see Secs. \ref{sec_ls} and
\ref{sec_euler}). However, the evolution of the perturbation is different. In the unstable regime $\omega=i\lambda$ with
$\lambda>0$, the perturbation grows exponentially rapidly as
$\delta\rho\sim e^{\lambda t}$. In the stable regime $\omega=-i\gamma$
with $\gamma>0$, the perturbation decreases exponentially rapidly as
$\delta\rho\sim e^{-\gamma t}$.

\subsection{Evolution of the spatial correlations}
\label{sec_cw}

In this section, we study the development of the spatial
correlations (at the order $1/N$) for a Brownian system in the
overdamped regime. For a homogeneous system, Eq. (\ref{s10}) for the
two-body correlation function reduces to
\begin{equation} \label{cw1}
{\partial h\over\partial t}={\partial\over\partial {\bf r}_{1}}\biggl\lbrack D_{*}{\partial h\over\partial {\bf r}_{1}}+\mu m^{2}{\partial u_{12}\over\partial {\bf r}_{1}}+\mu m\rho\int h({\bf r}_{2}-{\bf r}_{3},t){\partial u_{13}\over\partial {\bf r}_{1}}d^{D}{\bf r}_{3}\biggr\rbrack +(1\leftrightarrow 2),
\end{equation}
or, equivalently,
\begin{equation} \label{cw2}
{\partial h\over\partial t}=2D_{*}\Delta\biggl\lbrack h({\bf x},t)+\beta m^{2}u({\bf x})+\beta m\rho\int h({\bf y},t)u({\bf x}-{\bf y})d^{D}{\bf y}\biggr\rbrack,
\end{equation}
where ${\bf x}={\bf r}_{1}-{\bf r}_{2}$. Taking its Fourier transform, we get
\begin{equation} \label{cw3}
{\partial \hat{h}\over\partial t}+2D_{*}k^{2}\lbrack 1+(2\pi)^{D}\beta m\rho \hat{u}(k)\rbrack \hat{h}=-2D_{*}\beta m^{2}k^{2}\hat{u}(k).
\end{equation}
This equation is easily integrated in time to yield
\begin{equation} \label{cw4}
\hat{h}(k,t)=\hat{h}_{eq}(k)\biggl\lbrace 1-e^{-2D_{*}k^{2}\lbrack 1+(2\pi)^{D}\beta n m^{2}\hat{u}(k)\rbrack t}\biggr\rbrace,
\end{equation}
where $\hat{h}_{eq}(k)$ is the equilibrium value of the correlation
function (in Fourier space) given by Eq. (I-54) of Paper I. For the BMF
model (one mode), we find that
\begin{equation} \label{cw5}
{h}(\phi,t)={{\beta k/2\pi}\over 1-\beta/\beta_{c}}\biggl \lbrack 1-e^{-2D_{*}(1-\beta/\beta_{c})t}\biggr \rbrack\cos\phi.
\end{equation}
For $t\rightarrow +\infty$, the correlation function relaxes towards its equilibrium form (I-66). Note however  that the relaxation time diverges as we approach the critical point since $t_{relax}\sim (1-\beta/\beta_{c})^{-1}$.

\section{Generalized kinetic equations and effective
thermodynamics} \label{sec_gen}

\subsection{Generalized Kramers and Smoluchowski equations}
\label{sec_genk}

We shall introduce a class of stochastic processes leading to
generalized Fokker-Planck equations. These equations are associated
with an effective thermodynamical formalism in $\mu$-space. Let us first
consider the case of non-interacting Langevin particles described by
the stochastic process
\begin{equation}
\label{genk1} {d{\bf r}\over dt}=-\mu\nabla \Phi_{ext}({\bf
r})+\sqrt{K({\bf r},t)}{\bf R}(t),
\end{equation}
where ${\bf R}(t)$ is a white noise and $\Phi_{ext}({\bf r})$ an external
potential. Since the function in front of ${\bf R}(t)$ depends on
the position, the last term in Eq. (\ref{genk1}) can be interpreted as a
multiplicative noise. The corresponding
Fokker-Planck equation is
\begin{equation}
\label{genk2} {\partial \rho\over\partial t}={1\over 2}\Delta(K({\bf
r},t)\rho)+\mu\nabla (\rho\nabla\Phi_{ext}).
\end{equation}
When $K$ is constant, one recovers the usual Fokker-Planck
equation with a diffusion coefficient $D=K/2$. Recently, there was
some interest for the nonlinear Fokker-Planck equation
\begin{equation}
\label{genk3} {\partial \rho\over\partial
t}=\Delta(D\rho^{q})+\mu\nabla (\rho\nabla\Phi_{ext}),
\end{equation}
that arises in connection with Tsallis generalized thermodynamics,
see e.g. \cite{bukman}. Defining a generalized temperature $T=1/\beta$
through the Einstein-like relation $T=D/\mu$, we can rewrite the
foregoing equation in the form
\begin{equation}
\label{genk4} {\partial \rho\over\partial
t}=\nabla \lbrack D(\nabla\rho^{q}+\beta \rho \nabla\Phi_{ext})\rbrack.
\end{equation}
It is easy to verify that the nonlinear Fokker-Planck equation
decreases  the Lyapunov functional
\begin{equation}
\label{genk5}F[\rho]=\int \rho\Phi_{ext}\ d^{D}{\bf r}+{T\over
q-1}\int (\rho^{q}-\rho)d^{D}{\bf r},
\end{equation}
which can be interpreted as a free energy $F=E-TS$ associated with the
Tsallis entropy $S_{q}=-{1\over q-1}\int(\rho^{q}-\rho)d^{D}{\bf
r}$. Furthermore, the stationary solutions of this equation are given
by a $q$-distribution
\begin{equation}
\label{genk6}\rho=\biggl\lbrack \alpha-{\beta  (q-1)\over
q}\Phi_{ext}\biggr\rbrack^{1\over q-1},
\end{equation}
which minimizes the Tsallis free energy at fixed mass. In an attempt
to justify the nonlinear Fokker-Planck equation from a microscopic
model, Borland \cite{borland} proposed to consider the generalized stochastic
process
\begin{equation}
\label{genk7} {d{\bf r}\over dt}=-\mu\nabla \Phi_{ext}({\bf
r})+\sqrt{2D}\rho({\bf r},t)^{(q-1)/2}{\bf R}(t).
\end{equation}
The last term is a multiplicative noise which depends on ${\bf
r}$ and $t$ through the density $\rho({\bf r},t)$. Therefore,
there is a feedback from the macroscopic dynamics. For this
stochastic process, $K({\bf r},t)=2D\rho^{q-1}$ and the Fokker-Planck
equation (\ref{genk2}) takes the form (\ref{genk3}) where the diffusion coefficient
depends on the density.

In a previous work \cite{gt}, we remarked that the nonlinear
Fokker-Planck equation (\ref{genk4}) is a particular case of a larger
class of generalized Fokker-Planck equations associated with
generalized entropy functionals encompassing Tsallis entropy. Similar
observations have been made independently by Frank
\cite{frank} and Kaniadakis \cite{kaniadakis}. These equations can be written
\begin{equation}
\label{genk8} {\partial \rho\over\partial t}=\nabla\lbrace D\lbrack
\rho C''(\rho)\nabla\rho+\beta\rho\nabla \Phi_{ext}\rbrack\rbrace,
\end{equation}
where $C$ is a convex function, and they monotonically decrease the
functional
\begin{equation}
\label{genk9} F=\int\rho\Phi_{ext} d^{D}{\bf r}+T\int
C(\rho)d^{D}{\bf r},
\end{equation}
which can be interpreted as a generalized free energy. When
$C(\rho)=\rho\ln\rho$, we recover the usual Fokker-Planck equation
associated with the Boltzmann free energy and when
$C(\rho)=(\rho^{q}-\rho)/(q-1)$ we recover the nonlinear
Fokker-Planck equation associated with the Tsallis free
energy. Comparing Eq. (\ref{genk8}) with Eq. (\ref{genk2}) we find
that
\begin{equation}
\label{genk10} \mu=D\beta,\qquad {1\over 2}\nabla(K\rho)=D\rho
C''(\rho)\nabla\rho.
\end{equation}
The second equation is equivalent to
\begin{equation}
\label{genk11} {1\over 2}K\rho=\int^{\rho}D\rho C''(\rho)d\rho.
\end{equation}
Integrating by parts, we get
\begin{equation}
\label{genk12} {1\over 2}K\rho=D\lbrack \rho C'(\rho)-C(\rho)\rbrack,
\end{equation}
so that, finally,
\begin{equation}
\label{genk13} K({\bf r},t)=2D\rho\biggl\lbrack
{C(\rho)\over\rho}\biggr\rbrack'.
\end{equation}
Therefore, a stochastic process leading to the generalized
Fokker-Planck equation (\ref{genk8}) is given by
\begin{equation}
\label{genk14} {d{\bf r}\over dt}=-\mu\nabla \Phi_{ext}({\bf
r})+\sqrt{2D\rho\biggl\lbrack
{C(\rho)\over\rho}\biggr\rbrack'}{\bf R}(t),
\end{equation}
where ${\bf R}(t)$ is a white noise. We note in particular that the
strength of the stochastic force depends on the local density of
particles. When $C(\rho)=\rho\ln\rho$, we recover the usual Langevin
equation and when $C(\rho)=(\rho^{q}-\rho)/(q-1)$ we recover the
stochastic process considered by Borland \cite{borland}. The same
arguments can be developed in velocity space (taking into account the
inertia of the particles), see \cite{next2003}, leading to the generalized
Kramers equation
\begin{equation}
\label{genk15} {\partial f\over\partial t}+{\bf v}{\partial
f\over\partial {\bf r}}+\langle{\bf F}\rangle_{ext} {\partial
f\over\partial {\bf v}}={\partial\over\partial {\bf v}} \biggl
\lbrack D\biggl (fC''(f){\partial f\over\partial {\bf v}}+\beta
f{\bf v}\biggr )\biggr\rbrack.
\end{equation}

\subsection{Generalized non-local Kramers and Smoluchowski equations}
\label{sec_wq}

We shall now consider the generalization of the previous approach to
the case of particles in interaction.  We thus consider the $N$
coupled stochastic equations
\begin{eqnarray}
\label{wq1}
{d{\bf r}_{i}\over dt}={\bf v}_{i},
\end{eqnarray}
\begin{eqnarray}
\label{wq1b}
{d{\bf v}_{i}\over dt}=-\xi{\bf v}_{i}-\nabla_{i}U({\bf
r}_{1},...,{\bf r}_{N})+\sqrt{2Df_{i}\biggl\lbrack {C(f_{i})\over
f_{i}}\biggr\rbrack'}{\bf R}_{i}(t),
\end{eqnarray}
where $f_i=f({\bf r}_{i},{\bf v}_{i},t)$. We shall say that these
equations describe a gas of Langevin particles in interaction
\cite{anomalous}. We reserve the term of Brownian particles for usual
diffusion when $C(f)=f\ln f$ is the Boltzmann function (see
Sec. \ref{sec_k}). Using exactly the same steps as in
Sec. \ref{sec_k}, we can derive the generalized $N$-body Fokker-Planck
equation
\begin{equation}
\label{wq2} {\partial P_{N}\over\partial t}+\sum_{i=1}^{N}\biggl
({\bf v}_{i}{\partial P_{N}\over\partial {\bf r}_{i}}+{\bf
F}_{i}{\partial P_{N}\over\partial {\bf v}_{i}}\biggr
)=\sum_{i=1}^{N} {\partial\over\partial {\bf v}_{i}}\biggl\lbrack
{\partial\over\partial {\bf v}_{i}}\biggl (D f_{i}\biggl\lbrack
{C(f_{i})\over f_{i}}\biggr\rbrack' P_{N}\biggr )+\xi P_{N}{\bf
v}_{i}\biggr\rbrack.
\end{equation}
Implementing a mean-field approximation, we arrive at the generalized
non-local Kramers equation
\begin{equation}
\label{wq3} {\partial f\over\partial t}+{\bf v}{\partial
f\over\partial {\bf r}}+\langle{\bf F}\rangle {\partial
f\over\partial {\bf v}}={\partial\over\partial {\bf v}} \biggl
\lbrack D\biggl (fC''(f){\partial f\over\partial {\bf v}}+\beta
f{\bf v}\biggr )\biggr\rbrack,
\end{equation}
where $\langle {\bf F}\rangle$ is the mean-field force (\ref{k16}) and we have
set $\xi=D\beta$. This equation decreases the Lyapunov functional
\begin{equation}
\label{wq4} F[f]={1\over 2}\int f v^{2} d^{D}{\bf r} d^{D}{\bf
v}+ {1\over 2}\int \rho\Phi d^{D}{\bf r}+T\int C(f)d^{D}{\bf r}
d^{D}{\bf v},
\end{equation}
which can be interpreted as a generalized free energy.

Considering the strong friction limit
$\xi\rightarrow +\infty$, or the limit of large times $t\gg \xi^{-1}$,
we can derive a generalized Smoluchowski equation.  This is done in
\cite{gt} from the  moment equations of Eq. (\ref{wq3}) and in
\cite{lemou} by using a Chapman-Enskog expansion. The generalized
Smoluchowski equation can be written
\begin{equation}
\label{ge1} {\partial \rho\over\partial t}=\nabla \biggl \lbrack
{1\over\xi} (\nabla p+ \rho \nabla\Phi)\biggr\rbrack.
\end{equation}
The fluid is barotropic in the sense
that $p=p(\rho)$ where the equation of state is entirely specified by
the function $C(f)$. For the Boltzmann entropy, we recover the ordinary Smoluchowski equation (\ref{s13}) with $p=\rho {T\over m}$.  The generalized Smoluchowski equation (\ref{ge1}) decreases
the Lyapunov functional
\begin{equation}
\label{wq5}
F[\rho]=\int\rho\int_{0}^{\rho}{p(\rho')\over\rho'^{2}}d\rho'd^{D}{\bf
r}+{1\over 2}\int \rho\Phi d^{D}{\bf r},
\end{equation}
which can be interpreted as a generalized free energy. It can be
derived from the generalized free energy (\ref{wq4}) by using the
leading order expression of the velocity distribution $C'(f)=-\beta
\lbrack {v^{2}\over 2}+\lambda({\bf r},t)\rbrack+
O(1/\xi)$ to express $F[f]$ as a functional of $\rho$ (see
\cite{lemou} for details). Considering the dynamical stability of a spatially homogeneous solution of the generalized Smoluchowski equation (\ref{ge1}) with $\Phi=\rho * u$, we get the dispersion relation
\begin{equation}
\label{wq6}
i\omega\xi=c_{s}^{2}k^{2}+(2\pi)^{D}\hat{u}(k)k^{2}\rho,
\end{equation}
with $c_{s}^{2}=p'(\rho)$ which generalizes (\ref{s15}). For the damped Euler equations \cite{gt}, the term $i\omega\xi$ is replaced by $\omega(\omega+i\xi)$. For $\xi=0$ we recover (\ref{euler3}) and for $\xi\rightarrow +\infty$ we recover (\ref{wq6}).

We can also obtain a form of
generalized Smoluchowski equation by starting directly from the
coupled stochastic equations in physical space
\begin{equation}
\label{wq7} {d{\bf r}_{i}\over dt}=-\mu\nabla_{i}U({\bf
r}_{1},...,{\bf r}_{N})+\sqrt{2D_{*}\rho_{i}\biggl\lbrack
{C(\rho_{i})\over\rho_{i}}\biggr\rbrack'}{\bf R}_{i}(t),
\end{equation}
where $\rho_{i}\equiv \rho({\bf r}_{i},t)$ and ${\bf R}_{i}(t)$ is a
white noise acting independently on each particle. We note that these
equations cannot be obtained from Eqs. (\ref{wq1})-(\ref{wq1b}) by simply
neglecting the inertial term, as is done in the standard Brownian
case.  Applying steps similar to those of Sec. \ref{sec_s}, we obtain
the generalized $N$-body Fokker-Planck equation
\begin{equation}
\label{wq8} {\partial P_{N}\over\partial t}=\sum_{i=1}^{N}
{\partial\over\partial {\bf r}_{i}}\biggl\lbrack
{\partial\over\partial {\bf r}_{i}}\biggl (D_{*}\rho_{i}\biggl\lbrack
{C(\rho_{i})\over\rho_{i}}\biggr\rbrack' P_{N}\biggr )+\mu
P_{N}{\partial\over\partial {\bf r}_{i}}U({\bf r}_{1},...,{\bf
r}_{N})\biggr\rbrack.
\end{equation}
Implementing a mean-field approximation, we finally get the generalized non-local Smoluchowski equation
\begin{equation}
\label{wq9} {\partial \rho\over\partial t}= \nabla\lbrack D_{*} (\rho
C''(\rho)\nabla\rho+\beta \rho \nabla\Phi)\rbrack,
\end{equation}
where $\Phi$ is related to $\rho$ as in Eq. (\ref{k16}) and we have
set $\mu=D_{*}\beta$. This equation decreases the Lyapunov functional
\begin{equation} \label{wq10} F[\rho]={1\over 2}\int \rho\Phi
d^{D}{\bf r}+T\int C(\rho)d^{D}{\bf r}, 
\end{equation}
which can be interpreted as an effective free energy.

We note that the stationary solution of the generalized Fokker-Planck
equation (\ref{wq8}) is determined by the integro-differential equation
\begin{equation}
\label{wq11}
{\partial\over\partial {\bf r}_{i}}\biggl (D_{*}\rho_{i}\biggl\lbrack
{C(\rho_{i})\over\rho_{i}}\biggr\rbrack' P_{N}\biggr )+\mu
P_{N}{\partial\over\partial {\bf r}_{i}}U({\bf r}_{1},...,{\bf
r}_{N})={\bf 0},
\end{equation}
where $\rho({\bf r}_{i})=\int P_{N}d^{D}{\bf r}_{1}...d^{D}{\bf
r}_{i-1}d^{D}{\bf r}_{i+1}...d^{D}{\bf r}_{N}$. Contrary to the case
of Brownian particles studied in Sec. \ref{sec_s}, the equilibrium
$N$-body distribution does not seem to have a simple form.  In
particular, it does not seem to be possible to obtain a simple
generalization of the canonical distribution in
$\Gamma$-space by this approach. However, if we implement a mean-field
approximation, a notion of generalized thermodynamics emerges in
$\mu$-space, as we have seen. In this context, generalized free
energies are Lyapunov functionals associated with generalized
Fokker-Planck equations. The fact that we cannot obtain a simple form
of equilibrium distribution in $\Gamma$-space may suggest that the
generalized thermodynamical formalism developed in $\mu$-space is just {\it
effective}.

\subsection{Generalized Landau equation}
\label{sec_glw}

In \cite{physicaA}, we have proposed to consider a generalized class
of Landau equations of the form
\begin{eqnarray}
\label{glw1} {\partial f\over\partial t}=\pi
(2\pi)^{D}m{\partial\over\partial v^{\mu}}\int d^{D}{\bf
v}_{1}d^{D}{\bf k}k^{\mu}k^{\nu}{\hat{u}({\bf k})^{2}}\delta\lbrack
{\bf k} \cdot ({\bf v}-{\bf v}_{1})\rbrack f f_{1} \biggl (C''(f)
{\partial f\over\partial v^{\nu}}-C''(f_{1}){\partial
f_{1}\over\partial {v}_{1}^{\nu}}\biggr ),
\end{eqnarray}
where $C$ is a convex function. For $C=f\ln f$, we recover the
ordinary Landau equation (\ref{landau3}) as a particular case. As
shown in \cite{physicaA}, Eq. (\ref{glw1}) can be derived from the
generalized Boltzmann equation introduced by Kaniadakis
\cite{kaniadakis} in a weak deflexion limit. This equation conserves
mass and energy and increases the generalized entropy $S=-\int
C(f)d^{D}{\bf r}d^{D}{\bf v}$. This corresponds to a microcanonical
structure while the Fokker-Planck equation has a canonical structure
as it decreases the free energy at fixed mass and temperature.

In \cite{physicaA}, we have also considered a test particle approach and
showed the connection between the generalized Landau equation and the
generalized Fokker-Planck equation. Explicit expressions of the
diffusion coefficient have been obtained in $D=3$ for different
entropies $C(f)$.  We consider here the case $D=1$ and show that the
generalized Fokker-Planck equation  takes a relatively
simple form. First, the generalized Landau equation in $D=1$ reads
\begin{eqnarray}
\label{glw2} {\partial f\over\partial t}=K{\partial\over\partial v}\int d{v}_{1} \delta(v-v_{1})f f_{1} \biggl (C''(f){\partial f\over\partial v}-C''(f_{1}){\partial f_{1}\over\partial {v}_{1}}\biggr ),
\end{eqnarray}
where $K$ is given by Eq. (\ref{laudau11}). Once again, we see that
the collision term vanishes in $D=1$. In the test particle approach,
we must consider that $f_{1}$ is fixed to the distribution of the
bath, i.e. $f_{1}=f_{0}(v)$. This yields
\begin{eqnarray}
\label{glw3} {\partial f\over\partial t}={\partial\over\partial v} \biggl\lbrace D(v)\biggl \lbrack f C''(f){\partial f\over\partial v}-f {d C'(f_{0})\over d {v}} \biggr \rbrack\biggr\rbrace,
\end{eqnarray}
where $D(v)=Kf_{0}(v)$. This equation governs the evolution of the density probability $f({\bf v},t)$ of a single test particle in a bath of field particles with fixed distribution $f_{0}(v)$. We note that, for $t\rightarrow +\infty$, the test particle relaxes to the distribution of the bath, i.e. $f=f_{0}$.

\section{Conclusion}
\label{sec_con}

In this paper (Paper II), we have discussed the kinetic theory of
Hamiltonian and Brownian systems with long-range interactions. The
statistical equilibrium states of these systems have been considered
in Paper I. For Hamiltonian systems, in the $N\rightarrow +\infty$
limit, we get the Vlasov equation. This equation admits an infinity of
stationary solutions. One of them will be attained (on a
coarse-grained scale) as a result of a violent relaxation on a short
time scale of the order of the dynamical time $\sim t_{D}$. This
metaequilibrium state, or quasi-stationary state (QSS), which depends
on the initial conditions (due to the Casimir invariants) and on the
efficiency of the mixing process (ergodicity), is usually difficult to
predict
\cite{super}.  The statistical equilibrium state, due to the
development of correlations for finite $N$ systems, is reached on a
much longer timescale $t_{relax}\sim N^{\delta} t_{D}$ increasing with
the number of particles. For two and three-dimensional homogeneous
systems, we can develop the kinetic theory at order $1/N$ and get the
Lenard-Balescu equation.  This equation converges towards the
Maxwellian distribution establishing $\delta=1$. For inhomogeneous
gravitational systems, we get the Vlasov-Landau-Poisson equation when
the collisions are treated as local and collective effects are
neglected. This equation converges towards the mean-field
Maxwell-Boltzmann distribution. Its dynamical stability (with respect
to the Landau equation) coincides with the thermodynamical stability
criterion in the microcanonical ensemble (entropy maximum). Because of
logarithmic divergences of the diffusion coefficient, the relaxation
time scales as $t_{relax}\sim (N/\ln N) t_{D}$. Finally, for
one-dimensional systems (and 2D point vortices with monotonic angular
velocity profile), the collision term vanishes at the order $1/N$ so
that the evolution is due to higher order correlations. This implies
$\delta>1$ as observed by
\cite{yamaguchi} for the HMF model. For Brownian systems, in the  $N\rightarrow +\infty$ limit, we get the  non-local
Kramers and Smoluchowski equations. The mean-field Maxwell-Boltzmann
distribution is the only stationary solution of these equations. Its
dynamical stability (with respect to the Fokker-Planck equation)
coincides with the thermodynamical stability criterion in the
canonical ensemble (minimum of free energy).

We have also {\it formally} introduced generalized kinetic equations
and showed that they were associated with a generalized
thermodynamical framework in $\mu$ space. We have introduced these
equations (\ref{wq3}), (\ref{ge1}) and (\ref{wq9}) from a specific
class of stochastic processes (\ref{wq1})-(\ref{wq1b}) and (\ref{wq7})
but we believe that these equations can have interest in much more
general situations. They can be viewed as effective kinetic equations
attempting to describe complex media. They may be useful when we are
not in the strict conditions of applicability of standard kinetic
theories. For example, there are situations in which the two-body
distribution function cannot be factorized as a product of two
one-body distribution functions plus a small correction. In that case,
the system is not described by the ``ordinary'' kinetic equations
presented in Secs. \ref{sec_kin} and \ref{sec_gt}. On the
other hand, the dynamics of complex systems can be altered by
microscopic constraints (hidden constraints) that modify the form of
transition probabilities. Generalized kinetic equation can then be of
interest to describe (at least heuristically) these non-ideal
situations. One property of these generalized kinetic equations is to
exhibit anomalous diffusion (since the diffusion coefficient depends
on the density) and indeed, anomalous diffusion is observed in complex
media. In that case, this is associated with a complex geometrical
structure of phase space (e.g., multifractal in the case of the
Tsallis entropy). Note that anomalous diffusion can also be due to the
rapid decay of the diffusion coefficient with the velocity as in the
Fokker-Planck equation (\ref{isob2}). This is another, completely
independent, reason for anomalous diffusion
\cite{bd}. In that case, the Fokker-Planck equation is derived from
the pure Hamiltonian dynamics and there is no relation with
generalized thermodynamics. These two approaches apply to different
regimes or different systems.

\appendix

\section{The Landau equation and the temporal correlation function of the force without collective effects}
\label{sec_lt}

In this Appendix, we simplify the collision term appearing in the
kinetic equation (\ref{landau2}). For a homogeneous system, we need to
compute the memory function
\begin{eqnarray}
\label{lt1} M^{\mu\nu}=\int_{0}^{+\infty}dt\int d^{D}{\bf r}_{1} F^{\mu}(1\rightarrow 0,0) F^{\nu}(1\rightarrow 0,t).
\end{eqnarray}
The force created by particle $1$ on particle $0$ at time $t$ can be written
\begin{eqnarray}
\label{lt2}{\bf F}(1\rightarrow 0,t)=-i m\int {\bf k}\hat{u}({\bf k})e^{i{\bf k}({\bf r}(t)-{\bf r}_{1}(t))}d^{D}{\bf k}.
\end{eqnarray}
Making a linear trajectory approximation ${\bf r}(t)={\bf r}+{\bf v}t$ and ${\bf r}_{1}(t)={\bf r}_{1}+{\bf v}_{1}t$, we get
\begin{eqnarray}
\label{lt3}{\bf F}(1\rightarrow 0,t)=-i m\int {\bf k}\hat{u}({\bf k})e^{i{\bf k}({\bf x}+{\bf u}t)}d^{D}{\bf k},
\end{eqnarray}
where ${\bf x}={\bf r}-{\bf r}_{1}$ and ${\bf u}={\bf v}-{\bf v}_{1}$. Now, the memory function is easily calculated and yields
\begin{eqnarray}
\label{lt4} M^{\mu\nu}=\pi (2\pi)^{D}m^{2}\int k^{\mu}k^{\nu}\delta({\bf k}\cdot {\bf u})\hat{u}({\bf k})^{2}d^{D}{\bf k}.
\end{eqnarray}
Substituting this result in Eq. (\ref{landau2}), we get the Landau
equation (\ref{landau3}).

\section{The Lenard-Balescu equation and the temporal correlation function of the force}
\label{sec_leba}

In this Appendix, we give a short derivation of the Lenard-Balescu
equation. Its derivation is classical but we shall need some
intermediate steps in order to justify the expression (\ref{tc9}) of
the temporal correlation function. We follow the approach of
Padmanabhan \cite{paddy} but we consider an arbitrary potential of
interaction and we take into account collective effects. We start from
the Klimontovich equation
\begin{eqnarray}
\label{leba1}{\partial f_{d}\over \partial t}+{\bf v}\cdot {\partial f_{d}\over\partial {\bf r}}-\nabla\Phi_{d}\cdot {\partial f_{d}\over\partial {\bf v}}=0,
\end{eqnarray}
where $f_{d}({\bf r},{\bf v},t)=\sum_{i}m\delta({\bf r}-{\bf
r}_{i})\delta({\bf v}-{\bf v}_{i})$ is the exact discrete
distribution of particles and $\Phi_{d}({\bf r},t)$ is the potential
that they generate. Writing $f_{d}={f}+\delta f$ and
$\Phi_{d}={\Phi}+\delta\Phi$ and taking the average of Eq.
(\ref{leba1}), we get for a homogeneous distribution
\begin{eqnarray}
\label{leba2}{\partial f\over \partial t}={\partial\over\partial {\bf v}}\cdot \langle \delta f\nabla\delta\Phi\rangle.
\end{eqnarray}
Subtracting Eqs. (\ref{leba1}) and (\ref{leba2}) and neglecting the
quadratic terms (quasilinear approximation), we obtain
\begin{eqnarray}
\label{leba3}{\partial \delta f\over \partial t}+{\bf v}\cdot {\partial \delta f\over\partial {\bf r}}-\nabla \delta\Phi\cdot {\partial f\over\partial {\bf v}}=0.
\end{eqnarray}
It can be shown that the terms neglected are of order $1/N^{2}$. We can solve Eq. (\ref{leba3}) by Laplace-Fourier transform. This yields
\begin{eqnarray}
\label{leba4}(\omega-{\bf k}\cdot {\bf v})\delta\hat f=-{\bf k}\cdot {\partial f\over\partial {\bf v}}\delta\hat{\Phi}.
\end{eqnarray}
Therefore, the perturbed distribution function is given by
\begin{eqnarray}
\label{leba5}\delta\hat f({\bf k},\omega,{\bf v})=-{{\bf k}\cdot {\partial f\over\partial {\bf v}}\over \omega-{\bf k}\cdot {\bf v}}\delta\hat{\Phi}({\bf k},\omega)+\hat{g}({\bf k},\omega,{\bf v}),
\end{eqnarray}
where $\hat{g}$ is the general solution of $(\omega-{\bf k}\cdot {\bf
v})\hat{g}=0$.  For $\hat g=0$, after integration over the velocity
${\bf v}$, we obtain the dispersion relation $\epsilon({\bf
k},\omega)=0$ which arises in the linear stability analysis of the
Vlasov equation (see Sec. \ref{sec_ls}). The condition of marginal
stability corresponds to $\epsilon({\bf k},0)=0$. In the present
context, $\hat{g}$ is related to the discrete nature of the system,
i.e. the fact that the exact distribution function $f_{d}$ is a sum of
$\delta$-functions. Therefore, its expression is given by
\begin{eqnarray}
\label{leba6}\hat{g}({\bf k},\omega,{\bf v})={1\over (2\pi)^{D}}\sum_{i}m\delta({\bf v}-{\bf v}_{i})e^{i{\bf k}\cdot {\bf r}_{i}}\delta({\bf k}\cdot {\bf v}-\omega).
\end{eqnarray}
On the other hand, using the fact that the relation between the potential $\Phi$ and the distribution function $f$ is a convolution, we have in Fourier space
\begin{eqnarray}
\label{leba7}\delta\hat{\Phi}({\bf k},\omega)=(2\pi)^{D}\hat{u}({\bf k})\int \delta\hat{f}({\bf k},\omega,{\bf v}')d^{D}{\bf v}'.
\end{eqnarray}
Combining Eqs. (\ref{leba5}) and (\ref{leba7}) we obtain
\begin{eqnarray}
\label{leba8}\delta\hat{\Phi}({\bf k},\omega)=(2\pi)^{D}{\hat{u}({\bf k})\over \epsilon({\bf k},\omega)}\int \hat{g}({\bf k},\omega,{\bf v})d^{D}{\bf v},
\end{eqnarray}
where we have introduced the dielectric function (\ref{ls1}). It is now possible to compute the  collision current
\begin{eqnarray}
\label{leba9}\langle \delta f\nabla\delta\Phi\rangle=\int \langle \delta\hat{f}({\bf k},\omega,{\bf v})\delta\hat{\Phi}({\bf k}',\omega')\rangle i{\bf k}'e^{i({\bf k}+{\bf k}')\cdot {\bf r}}e^{-i(\omega+\omega')t}d^{D}{\bf k}d^{D}{\bf k}'d\omega d\omega'.
\end{eqnarray}
Using Eqs. (\ref{leba5}) and (\ref{leba8}), we get
\begin{eqnarray}
\label{leba10}  i{\bf k}'\langle \delta\hat{f}({\bf k},\omega,{\bf v})\delta\hat{\Phi}({\bf k}',\omega')\rangle=-(2\pi)^{2D}{\hat{u}({\bf k})\hat{u}({\bf k}')\over \epsilon({\bf k},\omega)\epsilon({\bf k}',\omega')}{i\over \omega-{\bf k}\cdot {\bf v}}{\bf k}'\biggl ({\bf k}\cdot {\partial f\over\partial {\bf v}}\biggr )\nonumber\\
\times \int \langle\hat{g}({\bf k},\omega,{\bf v}')\hat{g}({\bf k}',\omega',{\bf v}'')\rangle d^{D}{\bf v}'d^{D}{\bf v}''+(2\pi)^{D}{i\over \epsilon({\bf k}',\omega')}\hat{u}({\bf k}'){\bf k}'\int  \langle\hat{g}({\bf k},\omega,{\bf v})
\hat{g}({\bf k}',\omega',{\bf v}'')\rangle d^{D}{\bf v}''.\nonumber\\
\end{eqnarray}
From Eq. (\ref{leba6}), the correlation function is given by
\begin{eqnarray}
\label{leba11}\langle \hat{g}({\bf k},\omega,{\bf v}')\hat{g}({\bf k}',\omega',{\bf v}'')\rangle={m\over (2\pi)^{D}}f({\bf v}')\delta({\bf v}'-{\bf v}'')\delta({\bf k}\cdot {\bf v}'-\omega)\delta({\bf k}+{\bf k}')\delta(\omega+\omega').
\end{eqnarray}
Substituting these expressions in Eq. (\ref{leba9}) and using
\begin{eqnarray}
\label{leba12}\epsilon({\bf k},\omega)=1+(2\pi)^{D}\hat{u}({\bf k})\int \biggl \lbrack {P\over \omega-{\bf k}\cdot {\bf v}}-i\pi\delta(\omega-{\bf k}\cdot {\bf v})\biggr \rbrack {\bf k}\cdot {\partial f\over\partial {\bf v}} d^{D}{\bf v},
\end{eqnarray}
where $P$ denotes the principal part, we finally obtain the
Lenard-Balescu equation (\ref{bal1}). Moreover, from the above
expressions, it is easy to obtain the following expression for the
correlations of the potential
\begin{eqnarray}
\label{leba13}\langle \delta\hat{\Phi}({\bf k},\omega)\delta\hat{\Phi}({\bf k}',\omega')\rangle=m(2\pi)^{D}\delta({\bf k}+{\bf k}')\delta(\omega+\omega'){\hat{u}({\bf k})^{2}\over |\epsilon({\bf k},\omega)|^{2}}\int f({\bf v}')\delta({\bf k}\cdot {\bf v}'-\omega)d^{D}{\bf v}'.
\end{eqnarray}
Taking the inverse Fourier transform for the time variable, we obtain
\begin{eqnarray}
\label{leba14}\langle \delta\hat{\Phi}({\bf k},0)\delta\hat{\Phi}({\bf k}',t)\rangle=m(2\pi)^{D}\delta({\bf k}+{\bf k}')\hat{u}({\bf k})^{2}\int{1\over |\epsilon({\bf k},{\bf k}\cdot {\bf v}')|^{2}}e^{-i{\bf k}\cdot {\bf v}'t}f({\bf v}')d^{D}{\bf v}'.
\end{eqnarray}
Noting that the force acting on the particle at time $t$ is
\begin{eqnarray}
\label{leba15}{\bf F}(t)=-\int i{\bf k}\delta\hat{\Phi}({\bf k},t)e^{i{\bf k}\cdot {\bf v}t}d^{D}{\bf k},
\end{eqnarray}
we finally obtain the temporal correlation function of the force in the form
\begin{eqnarray}
\label{leba16}\langle F^{\mu}(0) F^{\nu}(t)\rangle=m(2\pi)^{D}\int k^{\mu}k^{\nu}{\hat{u}({\bf k})^{2}\over |\epsilon({\bf k},{\bf k}\cdot {\bf v}')|^{2}}e^{-i{\bf k}\cdot ({\bf v}-{\bf v}')t}f({\bf v}')d^{D}{\bf v}'d^{D}{\bf k}.
\end{eqnarray}
This expression generalizes Eq. (\ref{tc2}) in the case where collective effects are properly accounted for.

\end{document}